\pdfoutput=1
\documentclass[twocolumn, twocolappendix, numberedappendix]{aastex6}

\usepackage{caption}
\usepackage{graphicx}
\usepackage{latexsym}
\usepackage{amsfonts}
\usepackage{amsmath}
\usepackage{amssymb}
\usepackage{gensymb}
\usepackage{natbib}
\usepackage{url}
\usepackage{hyperref}
\usepackage{appendix}

\newcommand{\HI}{H{\sc\ i}}
\newcommand{\HII}{H{\sc\ ii}}
\newcommand{\SiIV}{Si{\sc\ iv}}
\newcommand{\CIV}{C{\sc\ iv}}
\newcommand{\OVI}{O{\sc\ vi}}

\newcommand{\SII}{S{\sc\ ii}}
\newcommand{\FeII}{Fe{\sc\ ii}}
\newcommand{\CII}{C{\sc\ ii}}
\newcommand{\PII}{P{\sc\ ii}}
\newcommand{\SiII}{Si{\sc\ ii}}
\newcommand{\SiIII}{Si{\sc\ iii}}
\newcommand{\MgII}{Mg{\sc\ ii}}
\newcommand{\AlIII}{Al{\sc\ iii}}

\newcommand{\kms}{km s$^{-1}$}
\newcommand{\NDH}{N_{\rm DH}^\perp}

\shorttitle{The MW's Hidden CGM \& COS-GAL}
\shortauthors{Y. Zheng et al. }

\begin{document}

\defcitealias{Savage09}{SW09}
\defcitealias{Wakker12}{W12}

\title{Revealing the Milky Way's Hidden Circumgalactic Medium with the \\Cosmic Origins Spectrograph Quasar Database for Galactic Absorption Lines}

\author{Y. Zheng$^{1, 2, 3}$, J. E. G. Peek$^{4, 5}$,  M. E. Putman$^3$, and J. K. Werk$^6$}
\affil{$^1$ Department of Astronomy, University of California, Berkeley, CA 94720, USA; yongzheng@berkeley.edu \\
       $^2$ Miller Institute for Basic Research in Science, University of California, Berkeley, CA 94720, USA\\
       $^3$ Department of Astronomy, Columbia University, New York, NY 10027, USA \\
       $^4$ Space Telescope Science Institute, 3700 San Martin Dr, Baltimore, MD 21218, USA \\
       $^5$ Department of Physics \& Astronomy, Johns Hopkins University, Baltimore, MD 21218, USA\\
       $^6$ Department of Astronomy, University of Washington, Seattle, WA 98195-1580, USA}

\begin{abstract}

Every quasar (quasi-stellar object; QSO) spectrum contains absorption-line signatures from the interstellar medium, disk-halo interface, and circumgalactic medium (CGM) of the Milky Way (MW). We analyze {\it Hubble Space
Telescope}/Cosmic Origins Spectrograph (COS) spectra of 132 QSOs to study the significance and origin of \SiIV\ absorption at $|v_{\rm LSR}|\leq100$ \kms\ in the Galactic halo. The gas in the north predominantly falls in at $-50\lesssim v_{\rm LSR}\lesssim 0$ \kms, whereas in the south, no such pattern is observed. The \SiIV\ column density has an average and a standard deviation of $\langle N_{\rm SiIV}\rangle=(3.8\pm1.4)\times10^{13}$ cm$^{-2}$. At $|b|\gtrsim30\degree$, $N_{\rm SiIV}$ does not significantly correlate with $b$, which cannot be explained by a commonly adopted flat-slab geometry. We propose a two-component model to reconstruct the $N_{\rm SiIV}-b$ distribution: a plane-parallel component $N_{\rm DH}^{\perp}$ to account for the MW's disk-halo interface and a global component $N_{\rm G}$ to reproduce the weak dependence on $b$. We find $N_{\rm DH}^{\perp}=1.3^{+4.7}_{-0.7}\times10^{12}$ cm$^{-2}$ and $N_{\rm G}=(3.4\pm0.3)\times10^{13}$ cm$^{-2}$ on the basis of Bayesian analyses and block bootstrapping. The global component is most likely to have a Galactic origin, although its exact location is uncertain. If it were associated with the MW's CGM, we would find $M_{\rm gas, all}\gtrsim4.7\times10^9\ M_{\odot} (\frac{C_f}{1})(\frac{R}{75\ {\rm kpc}})^2 (\frac{f_{\rm SiIV}}{0.3})^{-1}(\frac{Z}{0.3\ Z_{\odot}})^{-1}$ for the cool gas at all velocities in the Galactic halo. Our analyses show that there is likely a considerable amount of gas at $|v_{\rm LSR}|\leq100$ km s$^{-1}$ hidden in the MW's CGM. Along with this work, we make our QSO dataset publicly available as the COS Quasar Database for Galactic Absorption Lines (COS-GAL).


\end{abstract}


\section{Introduction}
\label{sec1}

The $\Lambda$CDM cosmology predicts that 16\%\footnote{$\Omega_b/\Omega_m=16$\%, $\Omega_b=0.0486$, $\Omega_m=0.308$; \citealt{Planck16}.} of the matter masses are in the form of baryons; however, observations of the low-redshift universe have so far found a limited amount of baryons. For example, \cite{Behroozi10} suggested that only $\sim10-20\%$ of the predicted baryons are locked in stars for galaxies at $z<4$. \cite{Peeples11} found that, for $L^*$ galaxies, only $\sim5$\% of the predicted baryonic mass is in cold gas form (i.e., \HI\ and H$_2$). The baryon deficit is even more severe over intergalactic scales, where less than 10\%\ of the predicted baryons exist in collapsed forms of stars in galaxies and hot gas in groups and clusters \citep{Persic92, Fukugita98, Cen99, Bregman07, Shull14}. Recent years have seen emerging evidence suggesting that galaxies are embedded in massive multiphase plasma, the so-called circumgalactic medium (CGM). The existence of the CGM helps to alleviate the missing baryon problem because, generally speaking, $\sim10^{9-11}M_\odot$ of baryonic mass is estimated in the CGM for sub- to super-$\L^*$ galaxies at $z\lesssim 0.5$ \citep{Tumlinson13, Werk13, Werk14, Stocke13, Liang14, Lehner15, Burchett16, Stern16, Prochaska17, Keeney17, Bordoloi18, Zahedy18}.


One of the tools proven most useful for studying the CGM of galaxies is absorption-line observations toward background quasars (quasi-stellar objects; QSOs), providing rest-frame observations of multiphase CGM with weakly (e.g., \MgII, \SiII) or highly (e.g., \SiIV, \OVI) ionized metals. With statistically significant samples of QSO sightlines probing the CGM at various impact parameters, it has been found that dense gas is preferentially located at smaller impact parameters (e.g., \citealt{Werk13}; \citealt{Lehner15}; \citealt{Keeney17}). In particular, \cite{Werk14} suggested a power-law form of the metal surface density of the CGM as $N_{\rm Si}\sim R^{-0.8}$ cm$^{-2}$. For star-forming galaxies, \cite{Werk13} found that the covering fraction of \SiIII\ and \SiIV\ reaches $\sim95\%$ within $75$ kpc of host galaxies, but the value drops to $\sim65\%$ at $75\leq R\leq 160$ kpc.



\begin{figure*}[t]
\begin{center}
\includegraphics[width=0.95\textwidth, height=0.95\textheight]{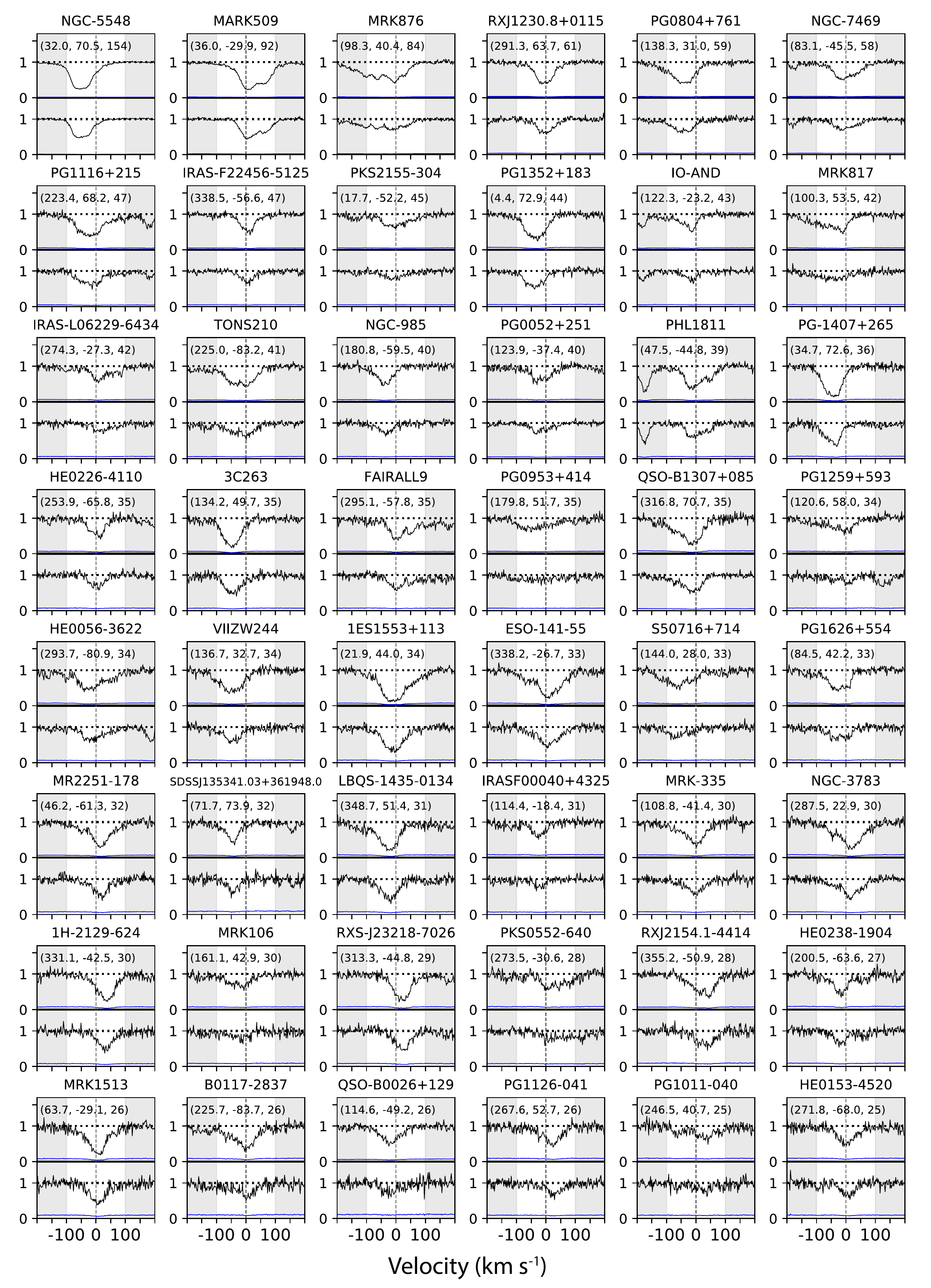} 
\caption{Continuum-normalized flux (black) and error (blue) of the \SiIV\ doublet for the 119 spectrally non-saturated QSOs (Q=0; see \S\ \ref{sec2.2}). Spectra are shown in original resolution. For each target, we show the \SiIV\ 1393/1402 lines in upper/low panels, respectively, and the numbers in parenthesis indicate $l$, $b$, and S/N per resolution element. Our analyses focus on the spectral region within $|v_{\rm LSR}|\leq100$ \kms, as indicated in white shades.}
\label{fig1}
\end{center}
\end{figure*}

\begin{figure*}[h]
\begin{center}
\includegraphics[width=0.95\textwidth, height=0.95\textheight]{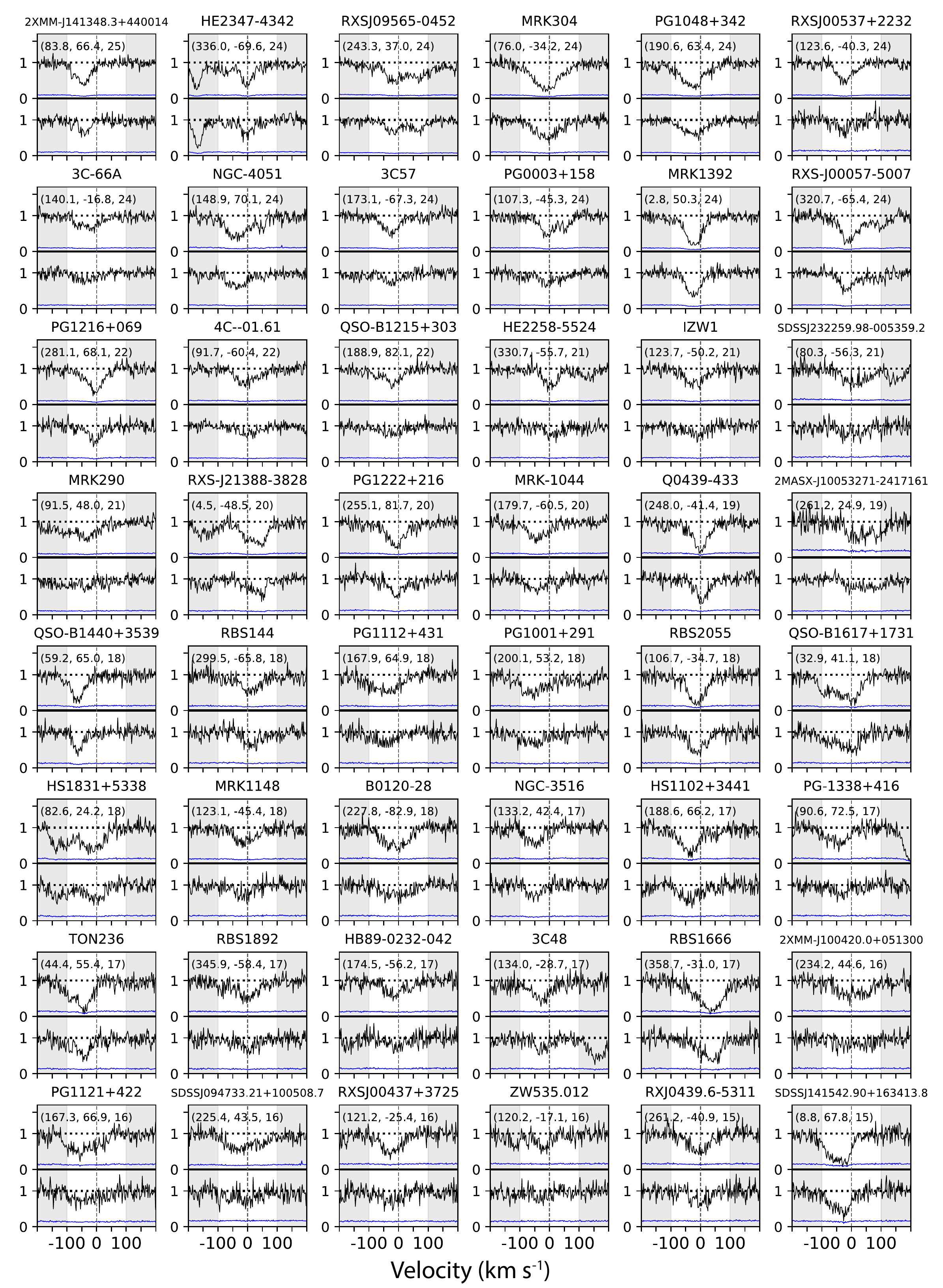}
\caption*{Figure \ref{fig1} continued. }
\end{center}
\end{figure*}

\begin{figure*}[t]
\begin{center}
\includegraphics[width=0.95\textwidth]{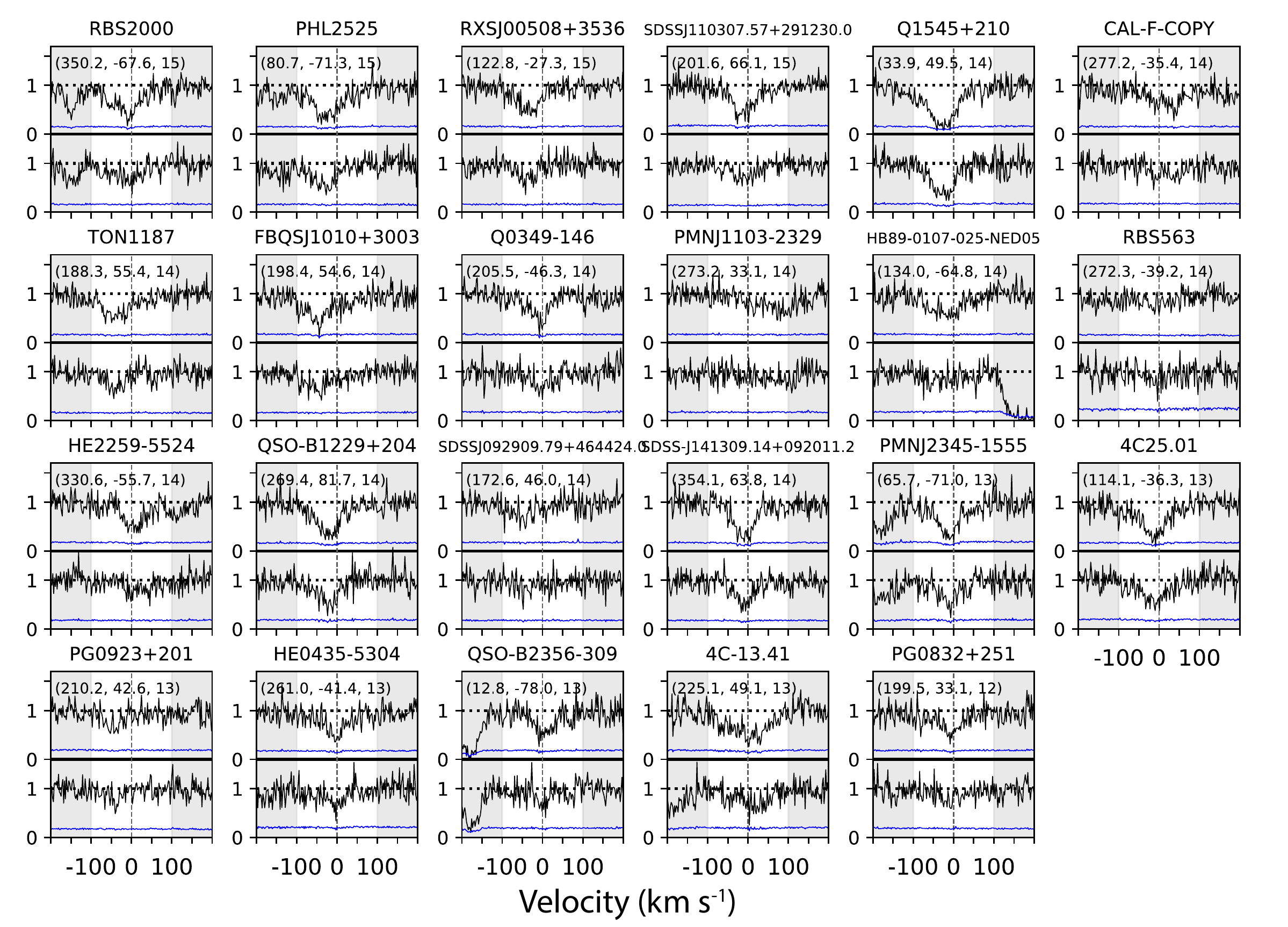}
\caption*{Figure \ref{fig1} continued. }
\end{center}
\end{figure*}

\begin{figure*}[t]
\begin{center}
\includegraphics[width=0.95\textwidth]{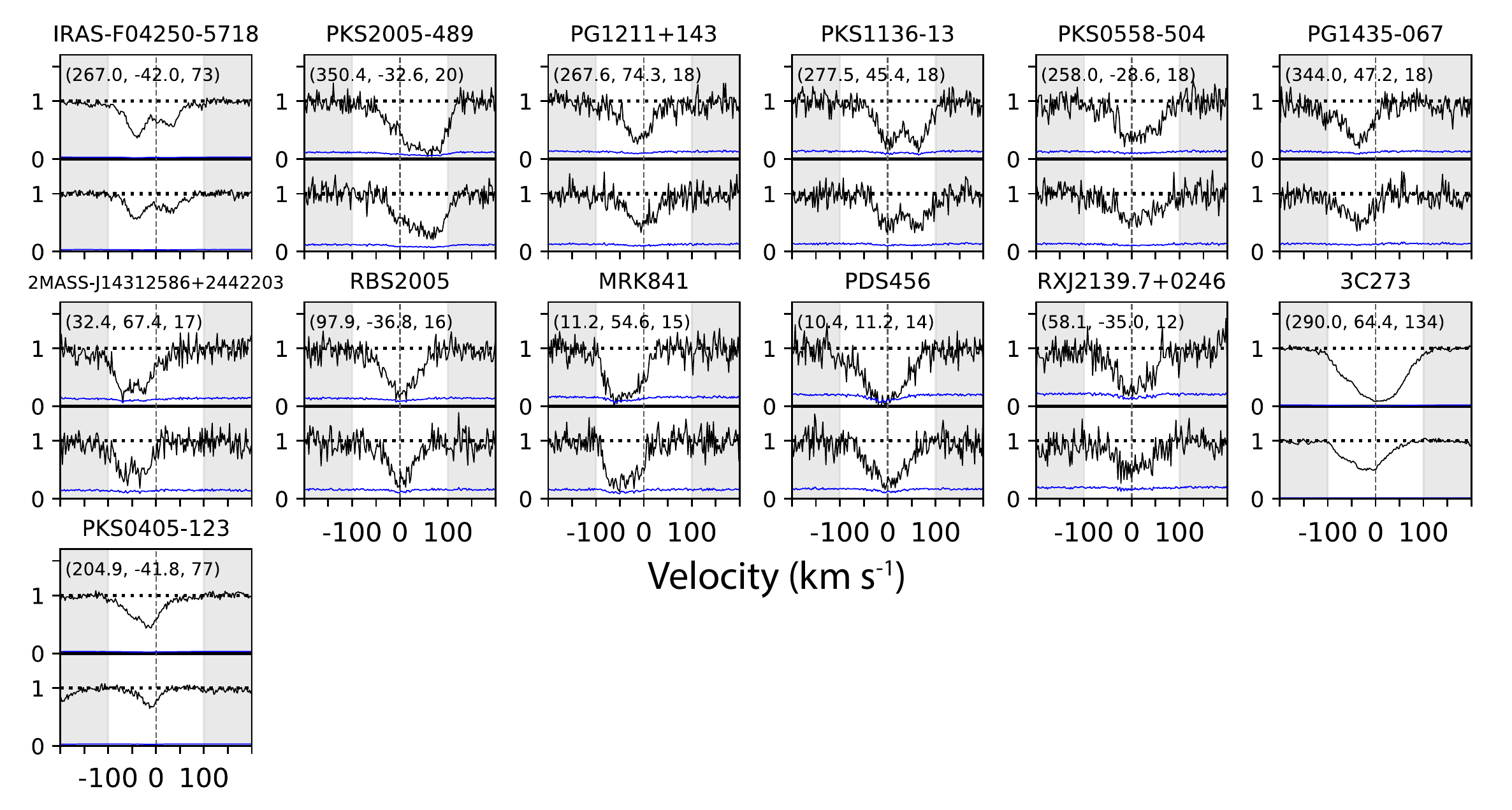}
\caption{Same as Figure \ref{fig1}, but for the 11 sightlines with unresolved saturation ($Q=1$; noted with ``S") and the two sightlines with abnormally strong \SiIV\ 1393 \AA\ profiles ($Q=-1$). See \S\ \ref{sec2.2} for more details. }
\label{fig2}
\end{center}
\end{figure*}

Inferred from the ubiquitous detection of CGM in extragalactic studies, our own Milky Way (MW) is most likely surrounded by a similar halo with abundant gas at various phases. The latest study of high-velocity gas in the MW halo using 270 QSO sightlines \citep{Richter17} showed that $77\%\pm6\%$ of the Galactic sky is covered by warm-ionized gas (e.g., \SiII, \SiIII, \CII, \CIV) moving at $|v_{\rm LSR}|=100-500$ \kms. They estimated that the high-velocity gas contributes $\sim3\times10^9\ M_\odot$ baryonic mass to the MW halo, 90\%\ of which comes from the Magellanic Stream \citep{Fox14}. The adopted high-velocity range is to avoid contamination of gas inside or near the Galactic disk; because as we sit inside the Galactic disk, every QSO spectrum unavoidably contains absorption-line signatures from the MW's interstellar medium (ISM), disk-halo interface, and the CGM. Therefore, velocity is often used as a proxy for distance: gas with $|v_{\rm LSR}|\gtrsim100$ \kms\ is selected to study those neutral/ionized high-velocity clouds at $\sim10-15$ kpc in the inner Galactic halo (e.g., \citealt{Sembach03, vanWoerden04, Wakker04, Shull09, Lehner12, Fox14}), whereas gas moving at $|v_{\rm LSR}|\lesssim 100$ \kms\ is typically deemed more nearby (see below). Hereafter we refer to the velocity range of $|v_{\rm LSR}|\lesssim 100$ \kms\ as the {\it low-intermediate} velocity gas, and refer to the range of $|v_{\rm LSR}|\gtrsim 100$ \kms\ as the {\it high}-velocity gas.




The low-intermediate velocity gas in the MW is studied using halo star sightlines and background QSO sightlines in the context of the disk-halo interface, which is modeled as a plane-parallel flat slab with a scale height of a few kiloparsecs above the Galactic plane (e.g., \citealt{Savage03, Savage09, Wakker12}). However, extragalactic CGM studies have shown that most ionic absorbers at a wide range of impact parameters are found to be gravitationally bound with centroid velocities of $|\delta_v| \lesssim200$ \kms\ from host galaxies' systemic velocities \citep{Werk13, Keeney17}.
In addition, \cite{Zheng15} pointed out that the MW's extended CGM is likely to host a similar amount of gas moving at both high and low-intermediate velocities, regardless of the gas phase, by making synthetic observations of a simulated MW-mass galaxy halo \citep{Joung12}.

Here we leverage the {\it Hubble Space Telescope} Spectroscopic Legacy Archive (HSLA; \citealt{Peeples17}) to revisit the ion distribution at low-intermediate velocities in the MW's disk-halo interface and CGM. Ions such as \CII, \SiII, and \SiIII, are so strong that they are often fully saturated and can only be used to study the gas at $|v_{\rm LSR}|>100$ \kms. For gas moving at $|v_{\rm LSR}|\leq100$ \kms, usually transition lines at high-ionization stages are preferred, such as \AlIII, \SiIV, \CIV, and \OVI\ (e.g., \citealt{Savage03, Savage09, Wakker12}). In this work, we focus on the \SiIV\ doublet at 1393/1402 \AA\ to study the warm-ionized gas at $|v_{\rm LSR}|\leq100$ \kms\ over the Galactic sky at $|b|\gtrsim30\degree$.

The article is organized as follows: \S\ \ref{sec2} introduces our data reduction of the {\it HST}/Cosmic Origins Spectrograph (COS) spectra, including continuum fitting, column-density and centroid-velocity calculations, and evaluation of line saturation. We make our continuum-normalized spectra publicly available as described in \S\ \ref{sec2.1}. In \S\ \ref{sec2.2}, we describe additional data reduction for the \SiIV\ $\lambda$1393/1402\AA\ doublet, which will be the focus of later analyses. In \S\ \ref{sec3} we study the all-sky distribution of the low-intermediate \SiIV\ absorbers. We compare our QSO measurements with a commonly adopted flat-slab model in \S\ \ref{sec4.1}, and propose a new two-component model in \S\ \ref{sec4.2}. In \S\ \ref{sec5}, we discuss the caveats of the flat-slab model and our model (\S\ \ref{sec5.1}), investigate the origin of the global component (\S\ \ref{sec5.2}), and conduct an order-of-magnitude mass estimate for the global component in the context of the MW's CGM (\S\ \ref{sec5.3}). We conclude in \S\ \ref{sec6}.

\section{Data}
\label{sec2}

\subsection{General Data Overview}
\label{sec2.1}

We obtain our QSO spectra from HSLA \citep{Peeples17}. The first release of HSLA (February 2017) delivers uniformly reduced and coadded {\it HST}/COS spectra with M-mode (medium-resolution/ G130M and G160M) and L-mode (low-resolution/ G140L) observations. For each target, the spectra from different programs, epochs, and gratings are coadded to improve the signal-to-noise ratio (S/N) and the wavelength coverage. Three community codes are considered for spectral coaddition: (1) {\it coadd\_x2d} \citep{Danforth10, Keeney12}, (2) {\it coscombine} \citep{Wakker15}, and (3) {\it counts\_coadd} \citep{Tumlinson13}. HSLA adopted the in-common steps among these codes: multiepoch spectra are shifted to the same wavelength grid by aligning strong interstellar lines that appear in every spectrum, and the wavelength grid is based on the values solved by the standard {\it calcos} pipeline. For the final products, HSLA arranges targets into different categories, such as {\it Solar System and Exoplanets}, {\it Stars} and {\it Galaxies and Clusters}. Detailed steps of spectral coaddition and target categories can be found in the HSLA document\footnote{http://www.stsci.edu/hst/cos/documents/isrs/ISR2017\_04.pdf} \citep{Peeples17}. In this work, we focus on the {\it QSOs, AGNs, and Seyferts} categories, which we will henceforth refer to as the HSLA QSO catalog.

For M-mode observations, the HSLA QSO catalog includes coadded spectra of 467 QSOs with G130M grating and 305 QSOs with G160M grating. In total, there are 511 QSOs, among which 261 QSOs were observed with both G130M and G160M. HSLA characterizes each coadded spectrum with an average S/N, defined as the mean of the S/N values per resolution element\footnote{G130M grating has six pixels per resolution (see the COS Data Handbook, Version 4, \citealt{Cos18}), therefore, S/N per resolution $\equiv$ S/N per pixel $\times\sqrt{6}$.} calculated over ten \AA\ windows every 1000 \AA\ starting from 1150 \AA\ \citep{Peeples17}. Our analyses focus on those QSOs with S/N $\geq5$, which includes 385 (246) QSOs with G130M (G160M) grating. In total, there are 401 QSOs, among which 230 QSOs have both G130M and G160M observations. For these 401 QSOs, we perform continuum normalization for a series of interstellar absorption lines and focus on the spectral region within $\pm1000$ \kms\ of the rest wavelength of each line. We use Linetools\footnote{https://github.com/linetools/linetools} package \citep{linetools} to conduct continuum normalization, which is an in-development open-source one-dimensional spectral analysis package.  Its continuum normalization function\footnote{http://linetools.readthedocs.io/en/latest/xspectrum1d.html} makes use of an Akima Spline to generate the continua from absorption-free regions near the lines of interest. An interactive interface is generated for each spectrum, and we visually inspect the suggested continua and make local correction to account for higher-order continuum variation if needed.

We compile the continuum-normalized interstellar line spectra for the 401 QSOs into a full dataset as the COS Quasar Database for Galactic Absorption Lines (COS-GAL). For each target, COS-GAL includes the normalized spectra of a number of interstellar lines, including \FeII\ $\lambda$1142/1143/1144/1608 \AA, \PII\ $\lambda$1152 \AA, \SII\ $\lambda$1250/1253/1259 \AA, \CII\ $\lambda$1334 \AA, \CIV\ $\lambda$1548/1550 \AA, \SiII\ $\lambda$1190/1193/1260/1526 \AA, \SiIII\ $\lambda$1206 \AA, and \SiIV\ $\lambda$1393/1402 \AA. We note that the continuum-normalized spectra provided in COS-GAL are specifically processed to study the absorption lines associated with the ionized gas of the MW. We do not conduct detailed analyses on higher-redshift intervening absorbers along the line of sight (LOS); users should exercise caution when inspecting these spectra for potential line contamination. In addition, the COS-GAL spectral lines are not ideal for studies of any broad features intrinsic to host QSOs. Users interested in these features should examine the original spectra instead.

In addition to the ion lines, COS-GAL also includes \HI\ 21cm emission lines extracted at the position of each QSO. We retrieve the \HI\ spectra from three \HI\ surveys at their native angular resolutions: GALFA-\HI\ (4$^\prime$; \citealt{Peek18}), HI4PI (16.2$^\prime$; \citealt{HI4PI}), and LAB (36$^\prime$;\citealt{Kalberla05}). We refer the reader to table 1 in \cite{HI4PI} for a comparison of the three surveys. Note that the COS spectra are observed with an aperture size of only 2.5$^{\prime\prime}$, which is much smaller than the radio beam sizes. Thus, we caution that the ionized gas studied by COS-GAL and the neutral gas extracted from these \HI\ surveys may not be entirely cospatial. We provide the \HI\ spectra at various resolutions to partially investigate this and refer the reader to a study by \cite{Wakker01} that examines the influence of different radio beam sizes on studies of high- and low-intermediate velocity clouds.

The COS-GAL dataset is made publicly available at \href{http://dx.doi.org/10.17909/T9N677}{[10.17909/T9N677]}. 
In the following sections, we focus on the \SiIV\ doublet $\lambda$1393/1402 \AA\ to study the warm-ionized gas moving at low-intermediate velocity in the MW.


\subsection{Data Reduction Focused on The \SiIV\ Doublet}
\label{sec2.2}

The \SiIV\ doublet is covered by G130M grating, and the COS-GAL dataset provides 385 QSOs in this grating with mean S/N$\geq$5. We examine the continuum-normalized \SiIV\ doublet for each QSO and discard 95 targets that are with either unmatched absorption components or very poor spectral quality. For the remaining 290 QSOs with reliable \SiIV\ absorption lines, we recalculate the S/N for the absorption-free continuum region between 1394 \AA\ and 1401\ \AA\ and select 132 high-quality QSO spectra with S/N per resolution $\geq12$. The \SiIV\ doublets of these 132 QSOs form the core sample of our following analyses.

Because the COS pipeline and HSLA adopt a heliocentric frame in their products, here we correct the spectral velocity to the LSR frame:
\begin{equation}
v_{\rm LSR}=v_{\rm hel}+9{\rm cos}(l){\rm cos}(b)+12{\rm sin}(l){\rm cos}(b)+7{\rm sin}(b)
\end{equation}
and present the continuum-normalized \SiIV\ doublets in Figures \ref{fig1} and \ref{fig2}. Separately, Figure \ref{fig1} shows the spectra of 119 QSOs classified as spectrally resolved ($Q=0$), whereas Figure \ref{fig2} includes 11 QSOs with unresolved saturation ($Q=1$) and two with uncertain contamination ($Q=-1$). The judgment of line saturation and the Q value is based on apparent column-density measurements, as described next.

To calculate the apparent \SiIV\ column densities, we adopt the apparent optical depth (AOD) method \citep{Savage91, Savage96}:
\begin{equation}
\begin{array}{l l}
N_{\lambda}(v)&=\frac{m_e c}{\pi e^2}\frac{-{\rm ln}[F_{n}(v)]}{f\lambda}\\
      &=3.768\times10^{14}\frac{-{\rm ln}[F_{n}(v)]}{f\lambda(\AA)} [{\rm atoms\ cm^{-2} (km\ s^{-1})^{-1}}], \\
\ \ \ \ N_{\lambda} &= \int_{v_-}^{v_+} N_{\lambda}(v){\rm d}v, \\
\ \ \ \ v_{\lambda} &= \int_{v_-}^{v_+} v N_{\lambda}(v){\rm d}v/\int_{v_-}^{v_+} N_{\lambda}(v) {\rm d}v,
\end{array}
\end{equation}
where $F_n(v)$ is the continuum-normalized spectrum, $\lambda$ is the line rest-frame wavelength, and $f$ is the oscillator strength. We adopt $\lambda$=1393.76/1402.77 \AA\ and $f$=0.513/0.254 for the \SiIV\ doublet from \cite{Morton03}. $N_\lambda$ is the total column density integrated over [$v_{\rm min}, v_{\rm max}$], and $v_\lambda$ is the flux-weighted centroid velocity. We adopt  $v_{-}=-100$ \kms\ and $v_{+}=+100$ \kms, respectively, as can be seen from Figures \ref{fig1} and \ref{fig2} that the majority of the MW \SiIV\ absorption features occur within this velocity range. The column densities and centroid velocities are tabulated in Table \ref{tb1}.

Among the 132 QSOs, we found 16 sightlines\footnote{They are with No. of 1, 2, 3, 4, 6, 7, 9, 14, 17, 19, 22, 24, 28, 36, 61, and 82 in Table \ref{tb1}.} in common with the {\it Far-Ultraviolet Spectroscopic Explorer} sightlines used in \citeauthor{Wakker12} (2012; henceforth \citetalias{Wakker12}). For these sightlines, we calculate the integrated column density over the same velocity ranges as those adopted in \citetalias{Wakker12}. We find that the mean differences between our COS values and theirs are 0.05 dex for \SiIV\ 1393\AA\ and 0.07 dex for \SiIV\ 1402\AA, indicating that our continuum fitting and column-density measurements are consistent with literature values.

\begin{figure}[t]
\begin{center}
\includegraphics[width=0.45\textwidth]{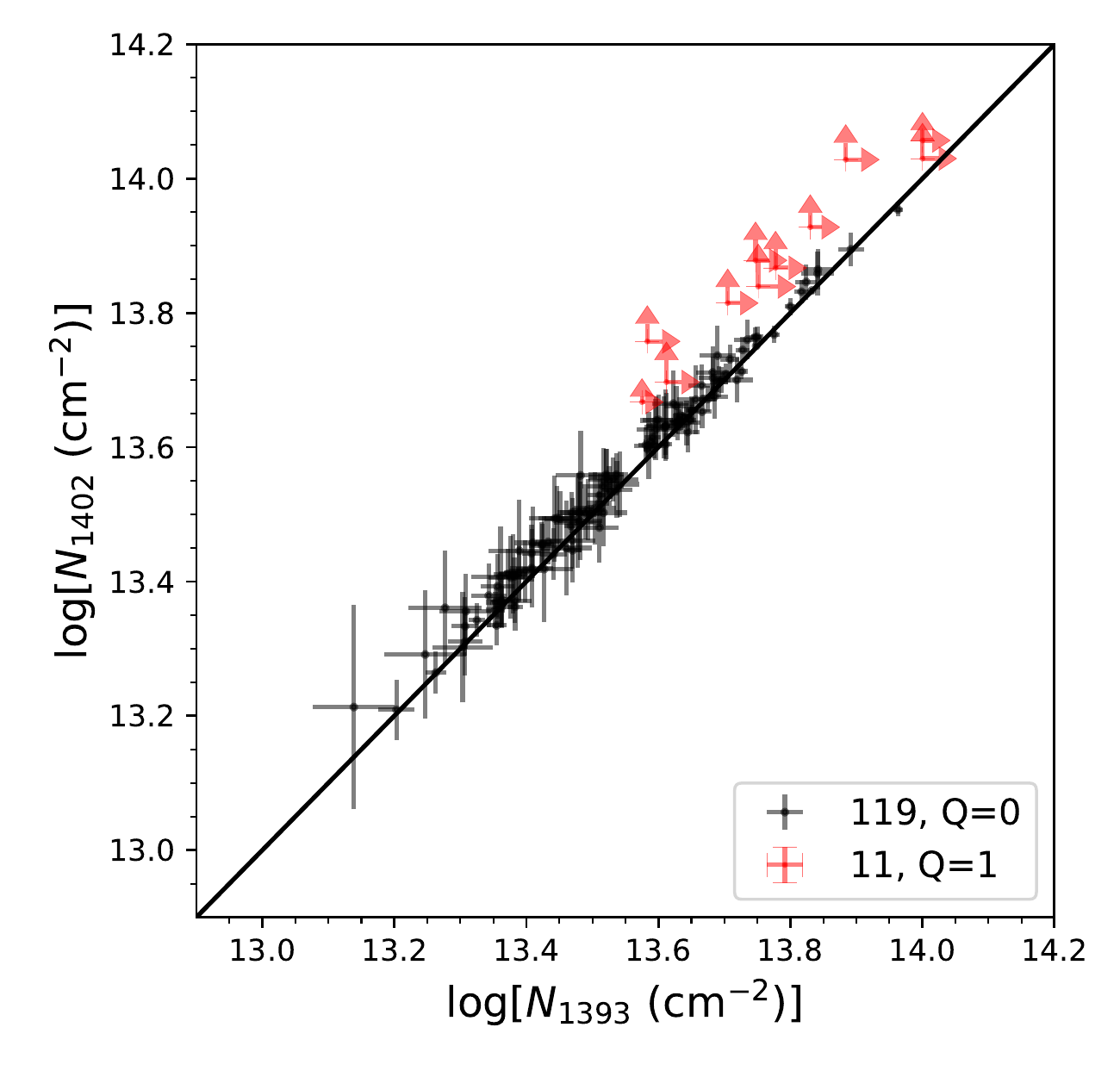}
\caption{Comparison of \SiIV\ doublet column-density measurements. In black are the 119 QSOs with spectrally resolved doublet profiles (Q=0), which meets the criterion in Equation \ref{eq3}. In red are the 11 sightlines with unresolved saturation (Q=1; see \S\ \ref{sec2.2}).}
\label{fig3}
\end{center}
\end{figure}

Even though the absorption-line profiles appear to be shallow, there might still exist spectrally unresolved saturation, as is pointed out by \cite{Savage91}. This effect is most significant when the full width half maximum (FWHM) of a strong absorption line is comparable to or smaller than the FWHM of the instrument's line-spread function. The unresolved saturation can also occur if multiple absorption components with various line widths coexist at similar velocity regions; the original absorption-line profiles end up being smeared into broader but shallower profiles. For ions with more than one transition line, they suggested an efficient way to identify unresolved saturation, which is to compare the $N_{\lambda}(v)$ profiles or the integrated $N_{\lambda}$ values between the transition lines. A spectrally well-resolved target should yield consistent $N_\lambda$ values among multiple absorption lines. Following this method, we define a QSO's \SiIV\ doublet as spectrally well-resolved (i.e., unsaturated) if the column densities of the doublet are consistent within their combined quadrature error:
\begin{equation}
\begin{array}{l l}
\delta {\rm log}N\equiv|{\rm log}N_{1393}-{\rm log}N_{1402}|\leq \sigma_c, \\
\sigma_c\equiv(\sigma_{1393}^2 + \sigma_{1402}^2)^{\frac{1}{2}},
\end{array}
\label{eq3}
\end{equation}
where $\sigma_c$ is the combined quadrature error, and $\sigma_{1393}$ and $\sigma_{1402}$ are the logarithmic errors of the corresponding ${\rm log}N_{1393}$ and ${\rm log}N_{1402}$ values as calculated from the spectra's error arrays. In addition, we require that an unsaturated \SiIV\ line should have its minimum normalized flux $\geq$ 0.1 because COS spectra are empirically known to be saturated when the normalized flux is lower than 0.1. The two requirements together find 119 (of 132) QSOs with unsaturated \SiIV\ doublets, which we classify as $Q=0$. The normalized doublet spectra are displayed in Figure \ref{fig1}, and their doublet column densities are shown as black symbols in Figure \ref{fig3}.



\begin{figure*}[t]
\begin{center}
\includegraphics[width=0.9\textwidth]{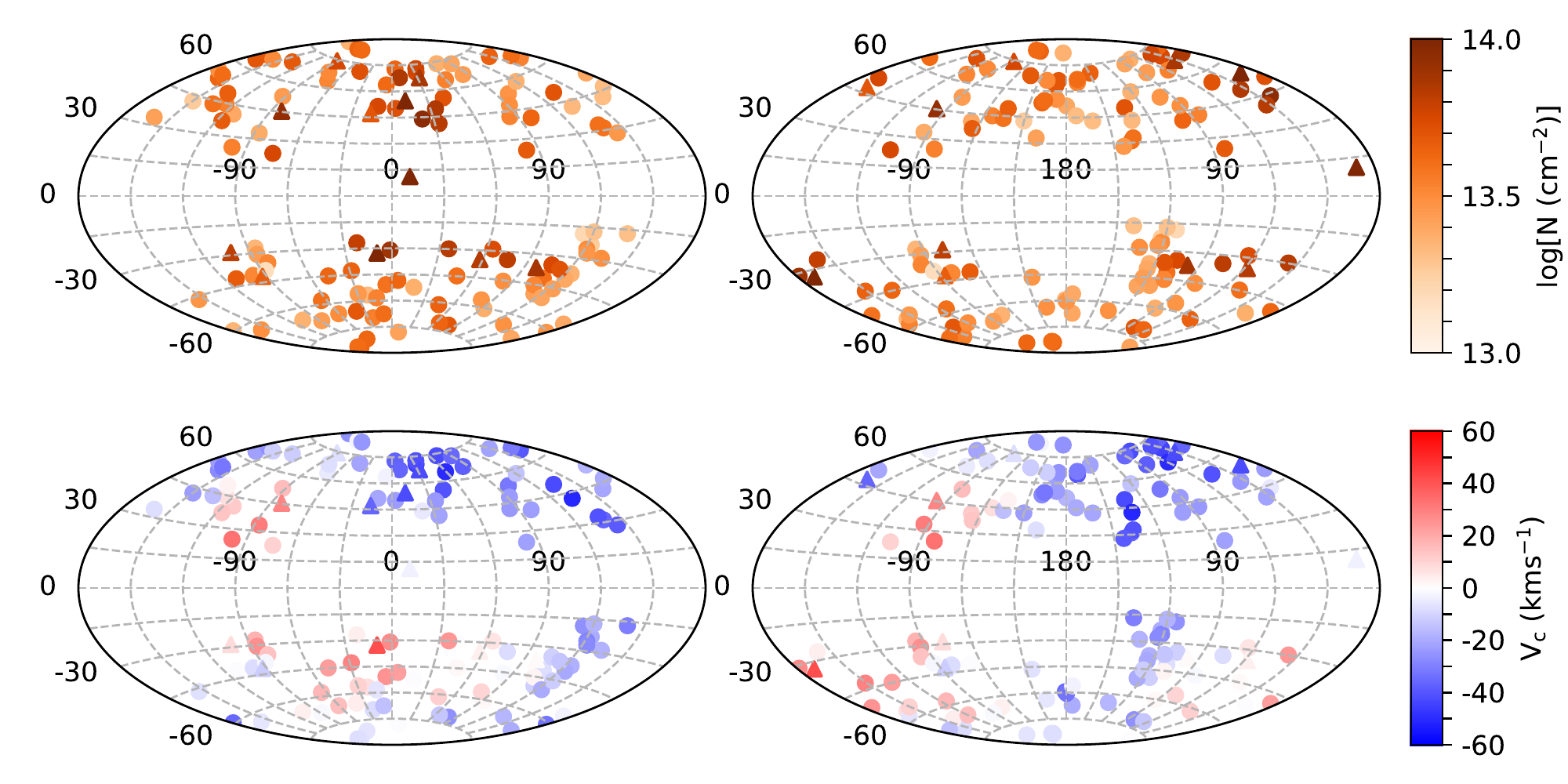}
\caption{The top two panels show the column-density distribution of the low-intermediate velocity \SiIV, with the left panel centered at the Galactic center and the right panel at anti-Galactic center. For spectrally non-saturated ($Q=0$) sightlines we show the mean column density of the doublet in dots; for spectrally saturated ($Q=1$) sightlines, we show the log$N_{1402}$ values as lower limits in upward triangles, given that \SiIV\ 1402\AA\ is weaker and, thus, is less saturated than 1393\AA. The bottom two panels show the distribution of the mean centroid velocity $v_{\rm c}$ (see \S\ \ref{sec3}). }
\label{fig4}
\end{center}
\end{figure*}

For the remaining QSOs, 11 of them either do not satisfy Equation \ref{eq3} or the minimum fluxes of the spectra are lower than 0.1. We flag these QSOs as spectrally saturated with Q=1, and show their doublet spectra in Figure \ref{fig2}. The column-density measurements for these QSOs are shown as red crosses in Figure \ref{fig3}, from which we find that saturation mostly occurs at high column densities. \cite{Savage91} provide column-density correction for unresolved saturated lines, which is based on a simulation of an isolated Gaussian component being smeared by instrument's line-spread function with different FWHM. As they pointed out, the corrections are most precise if the doublet's logarithm column-density difference is less than 0.05 dex. For targets with larger difference, the column-density correction itself could vary up to $0.1-0.3$ dex if the lines have multiple components. Given that our log$N$ measurement errors are typically $\sim0.05$ dex and that there could be multiple components in the \SiIV\ profiles as shown in Figure \ref{fig2}, it is unrealistic to apply column-density correction to these 11 saturated sightlines. We only list their column densities as lower limits in Table \ref{tb1}.

There remain two QSOs, 3C273 and PKS0405-123, which have ${\rm log}N_{1393}$ values more than $1\sigma_c$ higher than the ${\rm log}N_{1402}$ values. Their spectra are shown in the last two panels in Figure \ref{fig2}. For 3C273, \citetalias{Wakker12} suggests that the 1393 \AA\ line may suffer from blending with Ly$\alpha$. For PKS0405-123, we measure log$N_{1393}=13.46\pm0.01$, consistent with \citetalias{Wakker12}'s STIS$-$E140M value of log$N_{1393}=13.43\pm0.04$; note that the QSO is listed as PKS0405-12 in their work. However, for 1402 \AA\ we find log$N_{1402}=13.25\pm0.02$ while \citetalias{Wakker12} find $13.40\pm.07$. Neither line shows signs of a peculiar profile in our COS spectra (Figure \ref{fig2}) despite the log$N$ difference. As \citetalias{Wakker12} did not present the original spectra of PKS0405-123 in their work, we cannot conduct a close inspection of the line profiles to further compare results. We tag these two QSOs as Q=-1, and do not use them for future analyses. 




\section{The All-Sky Distribution of \SiIV\ with $|v_{\rm LSR}|\leq100$ \kms}
\label{sec3}

\begin{figure*}[t]
\begin{center}
\includegraphics[width=0.95\textwidth]{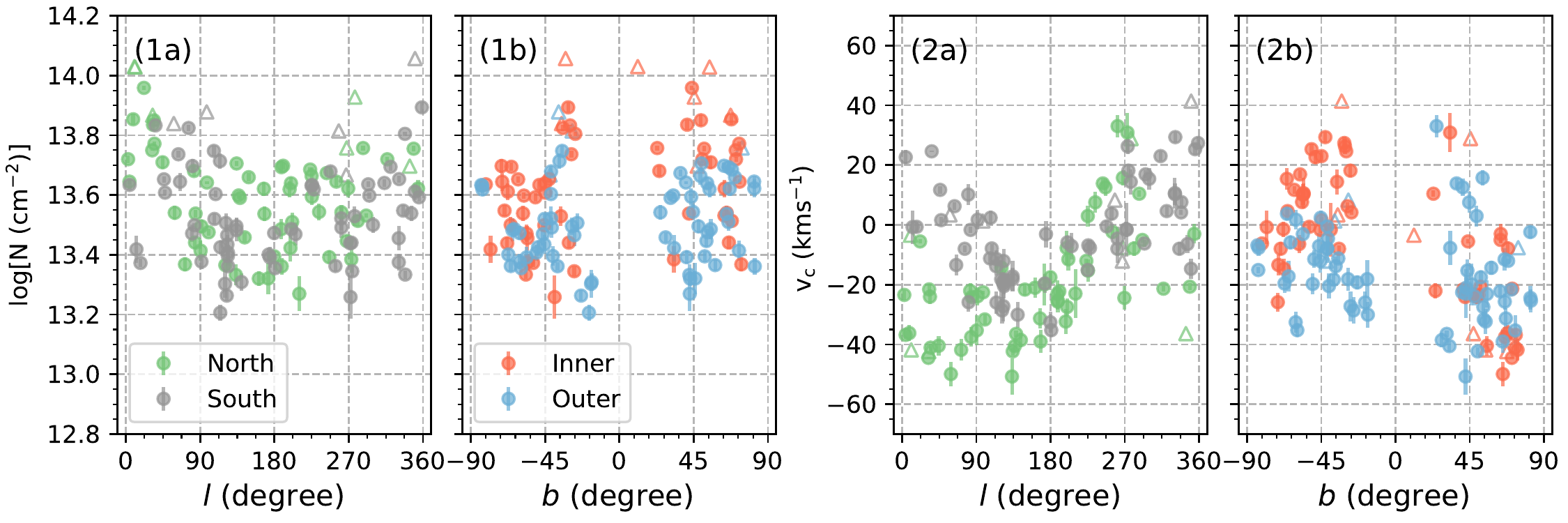}
\caption{Panels 1a/1b show the log$N$ distribution as functions of $l$/$b$, respectively. For spectrally non-saturated (Q=0) sightlines, we calculate the mean value of the doublet column density (dots); for spectrally saturated (Q=1) sightlines, we show the log$N_{1402}$ value as lower limits (upward triangles). In panel 1a, we color-code the data into North ($b>0\degree$, green) and South ($b\leq0\degree$, grey); in panel 1b, we split the data into Inner ($0\degree <l<90\degree$ or $270\degree<l <360\degree$, red) and Outer ($90\degree<l<270\degree$, blue) sections. Panels 2a/2b show the $v_c$ distribution as functions of $l$ and $b$; the symbols are color-coded the same way as those in panel 1a/1b, respectively. The centroid velocity is the mean of the doublet measurements for both the spectrally resolved and saturated sightlines.  }
\label{fig5}
\end{center}
\end{figure*}

In this section we study the column density (log$N$) and centroid velocity ($v_c$) distribution of \SiIV\ moving at $|v_{\rm LSR}|\leq100$ \kms\ over the Galactic sky, which we refer to as the {\it low-intermediate velocity} \SiIV. Overall the north and south are roughly evenly sampled with 62 and 68 QSO sightlines, respectively. In the top two panels of Figure \ref{fig4} and the left two panels of Figure \ref{fig5}, we show the mean doublet column densities of non-saturated (Q=0) sightlines in dots, and show the log$N_{\rm 1402}$ values of saturated (Q=1) sightlines in upper triangles to indicate the lower limits. For non-saturated (Q=0) sightlines, we find a mean column density and a standard deviation value of log$N=13.58\pm0.16$ dex [i.e., $N=(3.8\pm1.4)\times10^{13}$ cm$^{-2}$].

We find that the column densities of sightlines toward the Galactic center ($|l|\lesssim30\degree$) are generally $0.1-0.2$ dex higher than those in other regions, as shown in panel 1a in Figure \ref{fig5}. This could be due to material associated with the boundaries of the north and south Fermi Bubbles \citep{Ackermann14, Fox15, Bordoloi17}. Except for these sightlines, we do not see much difference between sightlines toward the inner ($l\leq90\degree$ or $l\geq90\degree$) or outer ($90\degree\leq l\leq 270\degree$) Galaxy (see Panel 1b in Figure \ref{fig5}). And, we find that the northern and southern sky has similar \SiIV\ column-density distribution, which is different from the observations of the low-intermediate velocity \OVI\  studied by \cite{Savage03}. There they found that the column density of \OVI\ at $b>45\degree$ is $\sim0.25$ dex higher than those in other regions. Note that the column densities of both the northern and southern sightlines show little dependence on $b$ as can be seen in panel 1b in Figure \ref{fig5}, which provides a critical constraint on the models that we discuss in \S\ \ref{sec4.1} and \S\ \ref{sec4.2}.





In the bottom two panels of Figure \ref{fig4} and the right two panels of Figure \ref{fig5}, we show the mean centroid velocities for both non-saturated (Q=0; dots) and saturated (Q=1; triangles) sightlines. We find that the blue-shifted pattern near $l\sim180\degree$ in panel 2a of Figure \ref{fig5} cannot be simply explained by co-rotation or halo lagging of ionized gas at the MW's disk-halo interface, which would otherwise show a clear segregation pattern of blue- and red-shifted velocity at $l=180\degree$ \citep{Wakker04}.
In panel 2b, most sightlines in the north show \SiIV\ moving toward us at $-50 \lesssim v \lesssim 0$ \kms. Since these sightlines are mostly at $b>30\degree$, their centroid velocities suffer only modest projection effects. The blue-shifted gas implies an excess of inflowing gas toward the disk; in the south, no obvious trend of gas inflows is found. This north-south dichotomy of velocity has also been observed in intermediate velocity clouds in \HI\ 21cm emission lines and metal absorption lines \citep[e.g.,][]{Wakker01, Schwarz04, Wakker04, Richter17}. We also note that the segregation of inner and outer Galaxy sightlines at $b<0\degree$ is likely an illustration artifact, as we do not find such sharp transition in the bottom right panel in Figure \ref{fig4} where smooth velocity field can be seen over the sky.

As we detect \SiIV\ absorbers along every sightline, the covering fraction (i.e., detection rate) of the low-intermediate velocity \SiIV\ at $|b|\gtrsim30\degree$ is $C_f$=100\%. This is similar to the results of ionized high-velocity gas \citep{Fox06, Shull09, Collins09, Lehner12, Richter17}. For example, \cite{Shull09} show that $81\%\pm5\%$ of the Galactic sky is covered by ionized high-velocity clouds according to a search of \SiIII\ absorbers in {\it HST}/COS and {\it FUSE} spectra of 37 AGNs, whereas \cite{Richter17} found a covering fraction of 74\%\ for high-velocity \SiIII\ using COS observations. 
With a different set of ions (e.g., \CII, \CIV, \SiIII, \SiIV), \cite{Lehner12} found a covering fraction of 68\% for gas moving at $|v|>90$ \kms\ at $|b|>20\degree$. We discuss the implication of the high covering fraction in \S\ \ref{sec5.3}.



\begin{figure*}[t]
\begin{center}
\includegraphics[width=0.6\textwidth]{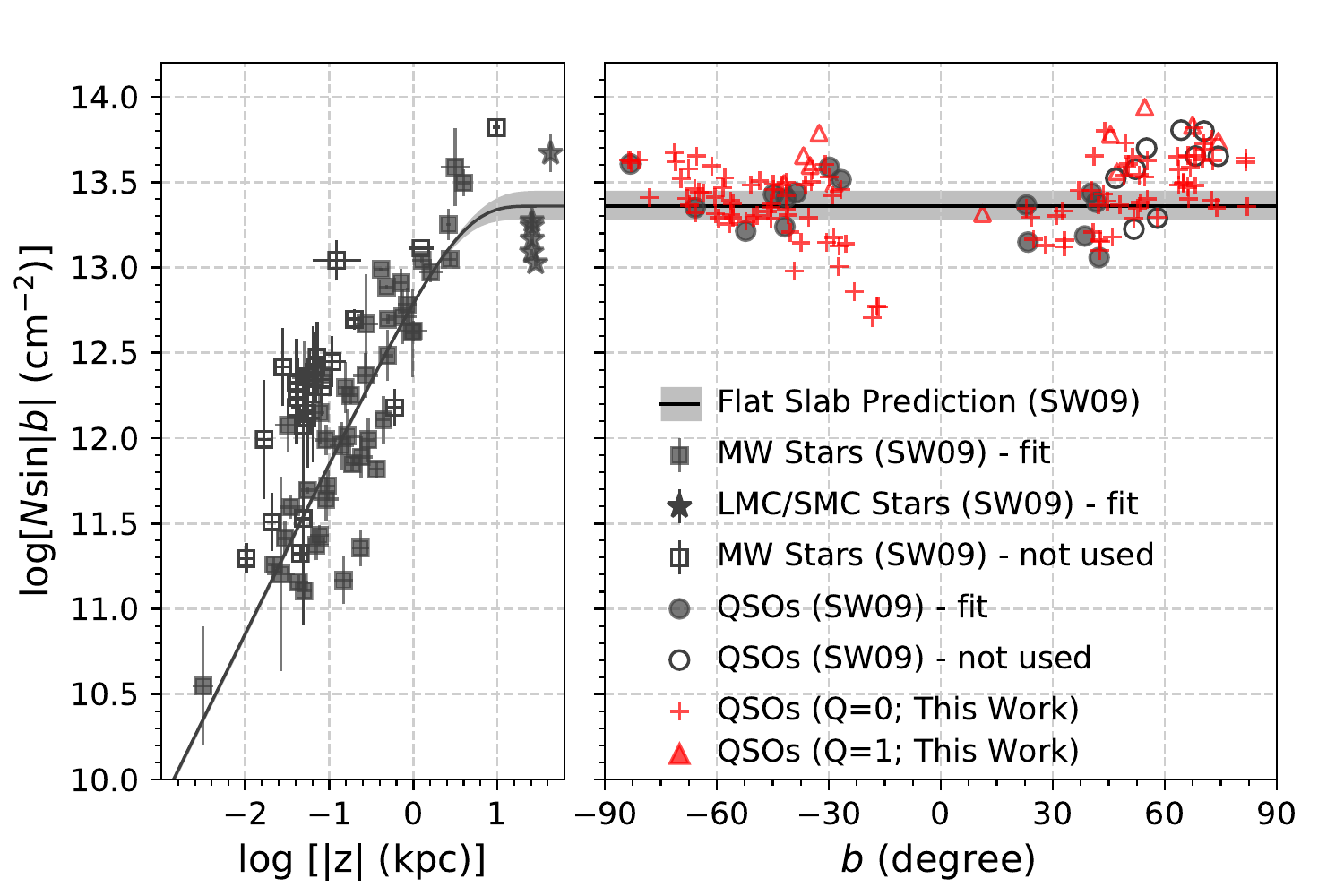}
\caption{The log($N$sin$|b|)$-log$|z|$ relation for stellar sightlines on the left panel and for QSOs on the right, respectively. In the left panel, the filled black symbols were used by \citetalias{Savage09} in their flat-slab model fitting,  including 44 MW halo stars (squares) and six LMC/SMC stellar sightlines (stars). Open symbols show those targets excluded from fitting, owing to sightlines intercepting specific ISM/halo structures. This plot is adopted from the \SiIV\ panel of figure 2 in \citetalias{Savage09}, although we do not include their upper/lower limit data. In the right panel, the red crosses/upward triangles are our \SiIV\ measurements of spectrally resolved (Q=0)/saturated (Q=1) sightlines, respectively; error bars are approximately equal to the size of the symbol. The black curve and gray shaded regions spanning from the left to the right panels show the predicted log$N_{\rm mod, \perp}$ and 1$\sigma$ values from \citetalias{Savage09}'s flat-slab model. Our QSO data show that log($N$sin$|b|$) increases as $|b|$ increases, which cannot be explained by the flat-slab geometry. See \S\ \ref{sec4.1} for more details. }
\label{fig6}
\end{center}
\end{figure*}

\section{Models for the \SiIV\ All-Sky Distribution}
\label{sec4}

Without distance constraints, it is difficult to trace the origins of the low-intermediate velocity \SiIV. In \S\ \ref{sec4.1}, we briefly review a commonly adopted plane-parallel flat-slab model which attributes the low-intermediate velocity \SiIV\ to the MW's disk-halo interface (e.g., \citealt{Savage90}; \citetalias{Savage09}; \citetalias{Wakker12}). Then in \S\ \ref{sec4.2}, we propose a better-fit two-component model which includes a global component in addition to the flat-slab geometry to account for the flat but scattered \SiIV\ column-density distribution as seen in panel 1b in Figure \ref{fig5}.

\subsection{A Commonly Adopted Flat-Slab Model}
\label{sec4.1}

The flat-slab model assumes a simple physical scenario that most of the low-intermediate velocity \SiIV\ is related to the MW's disk-halo interface (e.g., \citealt{Savage90}; \citealt{Savage09}, henceforth \citetalias{Savage09}; \citetalias{Wakker12}). The flat slab has an exponential density profile, $n(z)=n_0e^{-|z|/z_{\rm h}}$, extending above and below the Galactic disk. In this equation, $z$ is the vertical distance from the Galactic plane, $z_{\rm h}$ is the scale height of the gas, and $n_0$ is the midplane number density. $z_{\rm h}$ and $n_0$ can be constrained in the following way: for a star at height $z_{\rm s}$ from the Galactic plane, the modeled column density seen toward the star projected along the vertical direction is $N_{\rm mod, \perp}=\int_0^{z_s} n(z)dz$. $N_{\rm mod, \perp}$ can then be related to the observed column density toward the star as $N_{\rm mod, \perp}=N_{\rm obs}{\rm sin}|b|$, where $b$ is the star's Galactic latitude. Because the number density of the flat slab decreases dramatically above $z_{\rm h}$, the value of $N_{\rm mod, \perp}$ flattens as we probe background targets at vertical distances much larger than $z_{\rm h}$, such as stars in LMC/SMC and extragalactic QSO sightlines.

\citetalias{Savage09} modeled the distribution of warm-hot ionized gas moving at low-intermediate velocities over the Galactic sky using the flat-slab geometry. They collected the column densities of a number of ions (e.g., \HI, \AlIII, \SiIV, \CIV, \OVI) for 109 halo star sightlines and 30 QSO sightlines from the literature (e.g., \citealt{Jenkins78}; \citealt{Lehner07}; \citealt{Bowen08}; \citetalias{Wakker12}; see Table 2 of \citetalias{Savage09} for a complete list of references). Different authors have varied choices of velocity ranges over which the ion column densities were integrated, but the adopted ranges are all near $[-100, 100]$ \kms.
\citetalias{Savage09} examined the ${\rm log}(N_{\rm obs}{\rm sin}|b|)$-${\rm log}|z|$ relation, and found that ${\rm log}(N_{\rm obs}{\rm sin}|b|)$ increases with ${\rm log} |z|$ for most of their halo star sightlines, and the relation reaches a plateau for stars beyond certain heights. We reproduce \citetalias{Savage09}'s flat-slab experiment using their published stellar/QSO \SiIV\ data in Figure \ref{fig6}. In the left panel, the filled black symbols show the stellar sightlines used in their flat-slab parameter fitting. Sightlines noted as open symbols were not considered in their fitting because of either prominent \HII\ regions, X-ray sources, or supernova remnants along the lines of sight. They also excluded a halo star vZ 1182 at log$|z|\sim1.0$ and $b\sim79\degree$ because of the concern of contamination from the potential excess region near the north Galactic pole, as revealed by the \OVI\ all-sky distribution \citep{Savage03}. The right panel shows the QSO measurements as a function of Galactic latitude. The 15 QSOs shown in filled circles were used in their modelling, and there are ten sightlines at $b>45\degree$ (open circles) that were discarded to avoid the \OVI-enhanced region near the north Galactic pole \citep{Savage03}. The straight line and shaded area spanning from the left to the right panels in Figure \ref{fig6} show their model prediction of  the log$N_{\rm mod, \perp}-$log$|z|$ relation. They found a best-fit scale height of $z_h=3.2^{+1.0}_{-0.6}$ kpc and midplane density of $n_0=2.3\times10^{-9}$ ${\rm cm}^{-3}$ for the \SiIV\ flat slab.

Figure \ref{fig6} shows that the flat-slab model is able to reproduce the general trend of the projected \SiIV\ column densities seen toward halo stars. Since the model regards QSO sightlines as those with path lengths of infinity, the ${\rm log}(N_{\rm obs}{\rm sin}|b|)$ values from QSOs should always be ${\rm log}(N_{\rm obs}{\rm sin}|b|)=$ log($n_0z_h)=13.36$ (\citetalias{Savage09}). However, we find that this constant value cannot describe the all-sky QSO measurements. A Spearman test of ${\rm log}(N_{\rm obs}{\rm sin}|b|)-|b|$ for our data (red symbols in the right panel) finds a correlation coefficient of $r_{\rm s}=0.55$\footnote{A separate test for north- and south-only sightlines finds $r_{\rm s, n}= 0.53$ and $r_{\rm s, s}= 0.54$.}, implying ${\rm log}(N_{\rm obs}{\rm sin}|b|)$ is likely to increase monotonically with $|b|$, which indicates that a flat-slab geometry with an exponential density distribution is not adequate to explain the \SiIV\ content along QSO sightlines.

\begin{figure}[t]
\begin{center}
\includegraphics[width=0.45\textwidth]{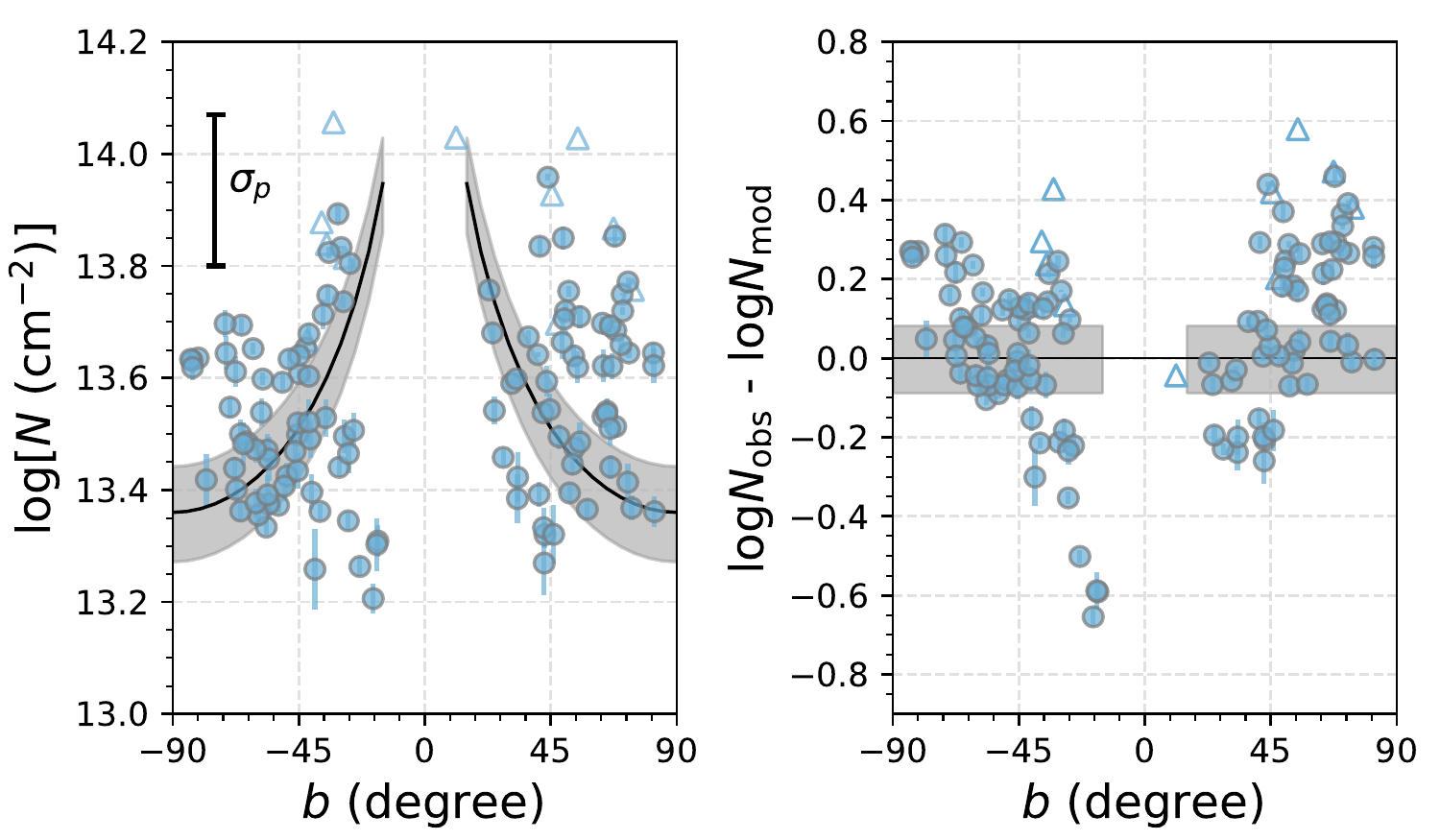}
\caption{Left: log$N$-$b$ relation for the low-intermediate velocity \SiIV. The data points are the same as those in Figure \ref{fig5}, but here we do not separate the inner or outer Galaxy data points because no significant trend related to these two segments is seen. Black curves show \citetalias{Savage09}'s flat-slab model prediction, and the gray shades are the $1\sigma$ ranges. The $\sigma_p$=0.27 dex in the top left is the patchiness parameter that \citetalias{Savage09} adopted, which we discuss in \S\ \ref{sec4.2}. Right: Residuals of the observed and model-predicted \SiIV\ column densities. A good model fit should yield a random distribution around 0.0 dex, however, the actual residuals show that the flat-slab model overpredicts the log$N$ value at low $b$ and underpredicts those at high $b$. }
\label{fig7}
\end{center}
\end{figure}

The mismatch between our QSO measurements and \citetalias{Savage09}'s flat-slab model prediction is further illustrated in Figure \ref{fig7}, where we directly compare the column-density values without sin$|b|$ projection. The black curves and gray shades show the log$N$ prediction from \citetalias{Savage09}'s best-fit flat-slab model, which suggests a strong dependence of log$N$ on $b$, i.e., QSO sightlines near polar regions have the shortest path lengths through the flat slab and thus should have the least \SiIV, while those near the Galactic plane have the longest paths and thus the most \SiIV. However, the observed log$N$ does not follow this predicted trend -- log$N$ seems to have a flat but scattered distribution regardless of sightlines' Galactic latitudes. The right panel shows the column-density residuals between the QSO data and \citetalias{Savage09}'s flat-slab model prediction. A good model fit should yield residuals that randomly scatter around 0.0 dex, which is not the case here. The flat-slab model over-predicts log$N$ at low $b$ and under-predicts those sightlines at high $b$, resulting in an increasing $\delta{\rm log}N$ at higher $b$.

We note that this mis-match between the measured log$N$ and flat-slab prediction has also been found in studies of Galactic \OVI\ (figure 12 of \citealt{Savage03}) and \CIV\ (figure 8 of \citetalias{Wakker12}) moving at low-intermediate velocities. In their work, \cite{Savage03} analyzed {\it FUSE} spectra of 100 extragalactic targets and pointed out that a symmetrical flat slab with an exponential density profile cannot accurately depict the distribution of \OVI\ moving at $|v|\lesssim100$ \kms. They also found that the \OVI\ centroid velocities do not fit into the scenario of a flat slab co-rotating with the Galactic disk. \CIV\ is likely to be a better proxy than \OVI\ for \SiIV\ given their comparable ionization potentials (33eV for \SiIV\ and 48eV for \CIV). In the flat-slab model, \citetalias{Savage09} found a scale height of $z_h=3.6$ kpc for \CIV, which is similar to \SiIV's scale height of $3.2$ kpc. \citetalias{Wakker12} show their log$N$ measurements of \CIV\ at low-intermediate velocity in their figure 8, overplotted with \citetalias{Savage09}'s flat-slab predictions. Their northern sightlines show a data-model mismatch similar to what we show in Figure \ref{fig7}. Their southern sightlines follow the general trend of the flat slab; however, because they have only 13 sightlines in the south, it is hard to clearly declare a 1/sin$|b|$ effect.


To summarize, the flat-slab geometry cannot fully describe the low-intermediate velocity \SiIV\ distribution seen along QSO sightlines. We proceed with a new model for the \SiIV\ distribution by adding in a global component to the flat-slab model in \S\ \ref{sec4.2}. Then we discuss the caveats of the flat-slab model and our model in \S\ \ref{sec5.1}.


\subsection{A Two-Component Model}
\label{sec4.2}

The spectrum of a QSO sightline going through the Galactic halo is likely to contain absorption-line information from the ISM, disk-halo interface, and the MW's CGM in a large volume simultaneously. The ISM is unlikely to contribute much to the \SiIV\ column density, as we discuss in \S\ \ref{sec5.2}. Here we construct a {\it two-component} model to evaluate the contribution of a disk-halo flat slab and a global component to the line-of-sight column densities. The disk-halo component takes into account the potential effect of Galactic latitude on a sightline's \SiIV\ column density: at low $b$, material at the disk-halo interface dominates the column density owing to a substantially increased path length through it (a factor of 2 for gas at $b = 30\degree$ compared to those at $b = 90\degree$). The global component reflects a uniform background to account for the weak $b$ dependence as shown in Figure \ref{fig5}. In the following, we describe the model setup and solve for the model parameters using a Bayesian analysis in \S\ \ref{sec4.2.1}. Our analysis follows the model-fitting discussion by \cite{Hogg10} and the Bayesian framework discussion in chapter 5.6 in \cite{Ivezic14}. Then in \S \ref{sec4.2.2} we resample the all-sky sightlines using block bootstrapping to estimate the influence of potential large-scale structures on our results.

\subsubsection{Bayesian Analysis on the Original Data Set}
\label{sec4.2.1}

We assume that the disk-halo and the global component components contribute to the (LOS) column density linearly:
\begin{equation}
N=\NDH\times \frac{1}{{\rm sin}|b|}+N_{\rm G},
\label{two_comp}
\end{equation}
where $\NDH$ is the column density of the disk-halo component projected onto the direction perpendicular to the Galactic plane. According to the flat-slab model (see \S\ \ref{sec4.1}), $\NDH$ should be a constant from sightline to sightline. $N_{\rm G}$ is the uniform global background, which is deemed invariant as well. 

For a sightline $k$ with column density $N_k$, we assume the measurement error $e_k$ is Gaussian. The probability distribution function (pdf) for a sightline with $N_k\pm e_k$ measurement given the two-component model is:
\begin{multline}
p_{\rm model}(N_k, e_k| \NDH, N_{\rm G}) \\
= \frac{1}{\sqrt{2\pi} e_k}{\rm exp}[-\frac{(N_k - N_{{\rm mod}, k})^2}{2 e_k^2}],
\label{p_mod}
\end{multline}
where $N_{{\rm mod}, k}\equiv \NDH/{\rm sin}|b_k|+N_{\rm G}$. In addition, we assume the data are drawn from an intrinsic Gaussian distribution with zero mean and variance of $\sigma_p^2$, where $\sigma_p$ is termed as the patchiness parameter. This assumption is to account for the scatter seen in the column-density measurements in Figure \ref{fig5} and \S\ \ref{sec3}, which is an order of magnitude larger than the measurement errors $\{e_k\}$. We discuss the physical meaning of $\sigma_p$ and its difference from the patchiness parameter used in \citetalias{Savage09} in \S\ \ref{sec5.1}. We can write down the probability of a sightline with column density $N_k$ drawn from an intrinsic $(0, \sigma_p^2)$ distribution:
\begin{equation}
p_{\rm scatter}(N_k|\sigma_p)=\frac{1}{\sqrt{2\pi} \sigma_p}{\rm exp}(-\frac{N_k^2}{2\sigma_p^2}).
\label{p_scat}
\end{equation}

We convolve Equations \ref{p_mod} and \ref{p_scat} to find the probability of a sightline $N_k$ following a linear relation (Equation \ref{two_comp}) with a Gaussian intrinsic scatter:
\begin{multline}
p(N_k, e_k| \NDH, N_{\rm G}, \sigma_p)=p_{\rm model}*p_{\rm scatter}\\
=\frac{1}{\sqrt{2\pi (e_k^2+\sigma_p^2)}}{\rm exp}[-\frac{(N_k - N_{{\rm mod}, k})^2}{2(e_k^2+\sigma_p^2)}].
\end{multline}
For brevity, we define $\theta\equiv(\NDH, N_{\rm G}, \sigma_p)$ as the parameter set in the following.

\begin{figure}[t]
\begin{center}
\includegraphics[width=0.5\textwidth]{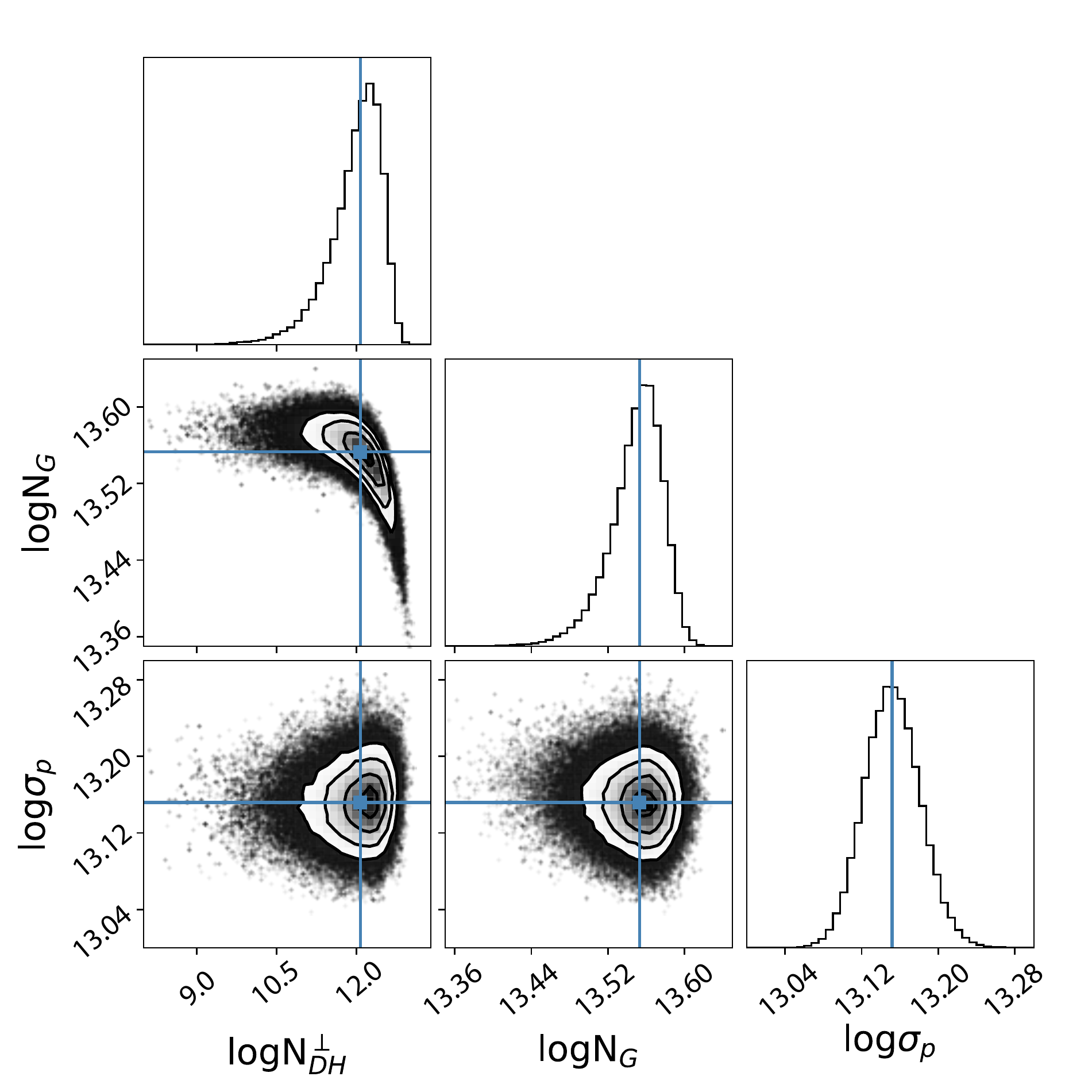}
\caption{The 1D and 2D marginalized posterior distributions for the parameters $\NDH$, $N_{\rm G}$, and $\sigma_p$. Note the large range differences on the axes for the three parameters.
The blue lines show the 50th percentile values for the parameters. This plot is generated with \href{https://github.com/dfm/corner.py}{corner.py} \citep{corner}. }
\label{fig8}
\end{center}
\end{figure}

Because our dataset includes K=119 QSO sightlines with non-saturated \SiIV\ doublets, we can calculate the likelihood of all these sightlines' LOS column densities being drawn from the two-component model with Gaussian intrinsic scatter as:
\begin{equation}
\mathcal{L}=\prod_{k=1}^K p(N_k, e_k| \theta).
\label{likelihood}
\end{equation}
A general approach to solve for the best-fit parameter set is to maximize $\mathcal{L}$ (or to minimize the logarithmic value, $\rm ln\ \mathcal{L}$). However, this maximum-likelihood method does not take into account a prior knowledge that the parameters ($\NDH$, $N_{\rm G}$, $\theta$) should have nonnegative values to yield physical meanings. The prior regarding this information can be written as:
\begin{equation}
        p(\theta|I)=\left\{
        \begin{array}{l l}
        const. & \NDH\geq0\ \&\ N_{\rm G}\geq0\ \&\ \sigma_p\geq0, \\
        0.0 & {\rm other\ cases},
        \end{array} \right.
\end{equation}
where the exact value of the constant is not important because we can renormalize over the posterior marginal pdf later. We can combine $p(\theta|I)$ with $\mathcal{L}$ to construct the posterior pdf for $\theta$:
\begin{equation}
p(\theta|\{N_k\}_{k=1}^K, I)=\frac{\mathcal{L} \times p(\theta|I)}{p(\{N_k\}_{k=1}^{K}|I)},
\end{equation}
The term in the denominator can be treated as a normalization constant so the integration for $p(\theta| \{N_k\}_{k=1}^K, I)$ over the parameter space will yield unity. The denominator term is trivial in our analysis, so in the following we will use $C$ to note any terms related to it. Conventionally, we calculate the logarithmic value of the posterior pdf:
\begin{equation}
\begin{array}{l l}
{\rm ln}[p(\theta|\{N_k\}_{k=1}^K, I)]&={\rm ln}\mathcal{L}+{\rm ln}[p(\theta|I)]+C\\
                                    &=\sum\limits_{k=1}^K {\rm ln}\frac{1}{\sqrt{2\pi(e_k^2+\sigma_p^2)}}\\
                                    &\ \ \ -\sum\limits_{k=1}^K\frac{(N_k-N_{{\rm mod}, k})^2}{2(e_k^2+\sigma_p^2)}\\
                                    &\ \ \ +{\rm ln}[p(\theta|I)]+C.
\end{array}
\label{post}
\end{equation}

To solve for ${\rm ln}[p(\theta|\{N_i\}_{i=1}^K, I)]$, we sample over the parameter space ($\NDH$, $N_{\rm G}$, $\sigma_p$) using the Markov chain Monte Carlo (MCMC) method with {\it emcee} \citep{emcee}, which is a Python implementation of the Affine Invariant MCMC Ensemble sampler \citep{mcmc}. In Figure \ref{fig8} we show the 1D and 2D marginalized distributions of the parameters. For the global component and the Gaussian intrinsic scatter, their posterior distributions show narrow profiles; we find the 50th percentile values at ${\rm log}N_{\rm G}=13.55$, and ${\rm log}\sigma_p=13.15$. For the disk-halo component, we find a broader distribution over a dynamic range of $\sim$(9, 13); the 50th percentile is ${\rm log}\NDH=12.07$. Figure \ref{fig8} shows that our modeling yields better constraints for $N_{\rm G}$ and $\sigma_p$ than for $\NDH$. This is likely to be due to the distribution of QSOs in our data set: because most QSOs are located at $|b|>30\degree$ (see Figure \ref{fig4}), the LOS column density is dominated by the global component, whereas the disk-halo component only becomes prominent at lower Galactic latitudes as a result of increased path length through the flat-slab medium (see \S\ \ref{sec5.1} for discussion on model caveats).

\subsubsection{Block Bootstrapping Over the Galactic Sky}
\label{sec4.2.2}

Our previous analysis makes the implicit assumption that the sightlines are independent and uniformly distributed, which is not a well-defined assumption, because the gaseous Galactic sky is covered by various structures, such as the large complexes moving at intermediate velocities in the northern sky (e.g., IV Arch and IV Spur; \citealt{Wakker04}) and the Fermi Bubbles. Although our data have sampled the north and south quite evenly (62 QSOs in the north and 68 in the south), it is possible that some sightlines are correlated at certain spatial scales. Here we divide the Galactic sky into eight segments: each segment is $90\degree\times90\degree$ in size, with the first segment at ($l$, $b$)$=$($0\degree$, $-90\degree$). By block bootstrapping these eight segments and rerunning the MCMC using the newly sampled datasets 10$^4$ times, we reduce the covariance due to very large structures.

\begin{figure}[t]
\begin{center}
\includegraphics[width=0.4\textwidth]{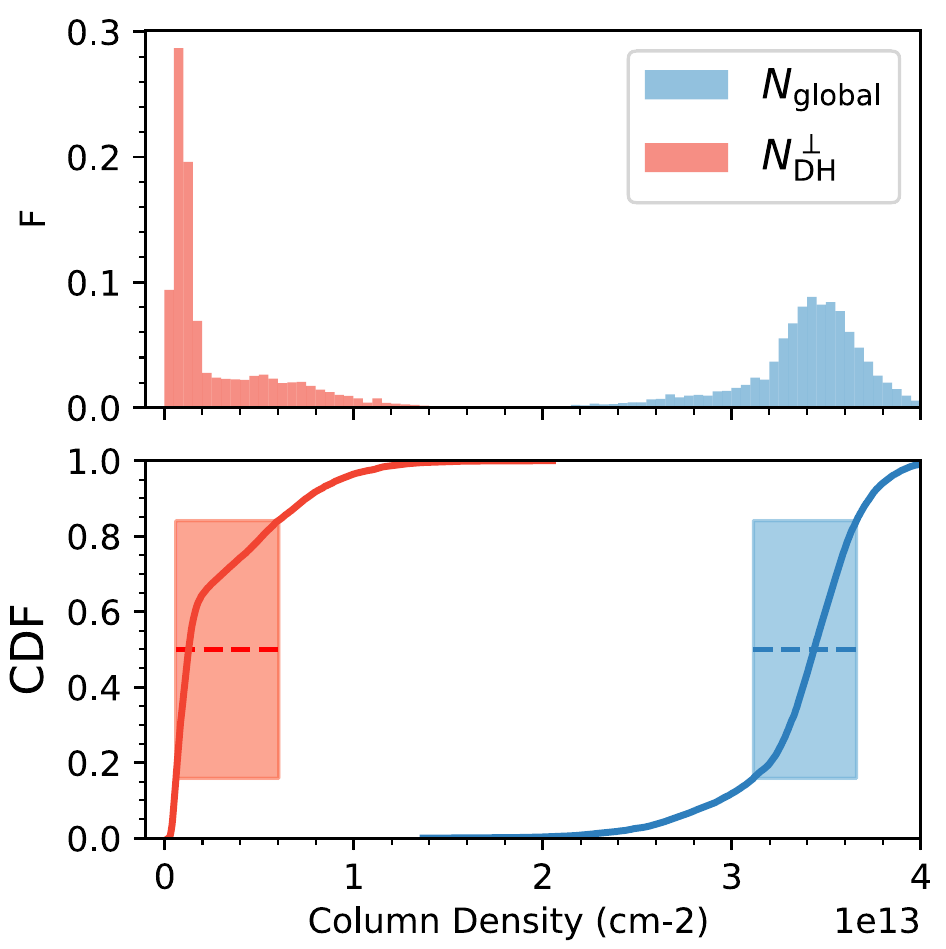}
\caption{Top: distribution of the 50th percentile values of $\NDH$ (red, left) and $N_{\rm G}$ (blue, right) from 10$^4$ runs of block bootstrapping. The data are binned in steps of $5\times10^{11}\ {\rm cm^{-2}}$. Bottom: The cumulative distribution function (CDF) for $\NDH$ and $N_{\rm G}$. The square patches show the 1$\sigma$ (68.2\%) credible ranges. }
\label{fig9}
\end{center}
\end{figure}

For each run, we repeat the MCMC process and calculate the 50th percentile values of ($\NDH$, $N_{\rm G}$, $\sigma_p$). We show the distribution of the 50th percentile values for $\NDH$ and $N_{\rm G}$ in Figure \ref{fig9}. In the top panel, the data are binned in steps of $5\times10^{11}\ {\rm cm^{-2}}$ for both $N_{\rm DH, \perp}$ and $N_{\rm G}$. The disk-halo component $N_{\rm DH, \perp}$ shows a sharp peak at low column-density values less than $2\times10^{12}$ cm$^{-2}$, whereas the global component $N_{\rm G}$ has a broad base with a peak at $\sim3.4\times10^{13}$ cm$^{-2}$. We show the cumulative distribution functions (CDFs) of $\NDH$ and $N_{\rm G}$ in the bottom panel. At the 68.2\% confidence level ($1\sigma$), we find $\NDH=1.3^{+4.7}_{-0.7}\times10^{12}$ cm$^{-2}$ and $N_{\rm G}=3.4^{+0.3}_{-0.3}\times10^{13}$ cm$^{-2}$. Similarly, we find the patchiness parameter $\sigma_p$ has a median value of $\sigma_p=1.4\times10^{13}$ cm$^{-2}$, consistent with the column-density scatter measured from the QSO data.

\begin{figure}[t]
\begin{center}
\includegraphics[width=0.48\textwidth]{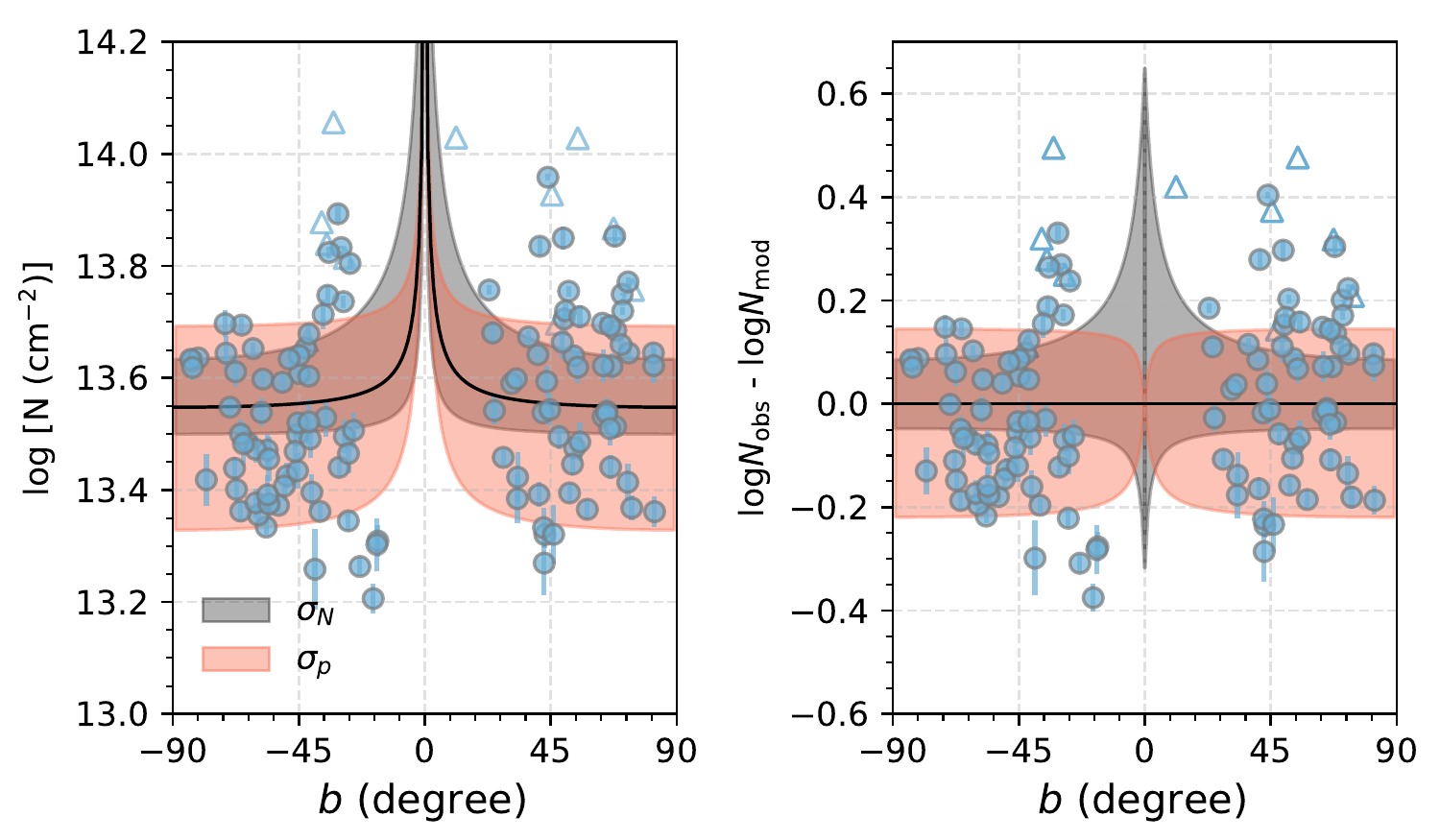}
\caption{Left: log$N$-$b$ distribution. The data points are the same as those in Figure \ref{fig7}. The black line shows the predicted column densities from our two-component model and the gray shades indicate the 68.2\% (1$\sigma$) confidence range as inferred from the CDF (Figure \ref{fig9}). The red shades show the patchiness parameters, which cover most of the data points, indicating that our model is able to account for the intrinsic scatter of the LOS column density. Right: column-density residuals of the observed and model-predicted \SiIV\ column densities. A Spearman test on the data-model residuals finds a correlation coefficient of $r_s$ = 0.13, implying that the abnormal trend of $\delta$log$N$ increasing with $b$ seen in Figure \ref{fig7} has now disappeared. }
\label{fig10}
\end{center}
\end{figure}


We compare our two-component best fit with the QSO measurements in Figure \ref{fig10}. At low Galactic latitudes where the LOS path goes deeply through the flat slab, the disk-halo component contributes to the majority of the column density; while at high Galactic latitudes, the global component dominates the LOS column densities. In addition, as highlighted in red shades, the Gaussian-form patchiness parameter we include in the model is able to recover most of the intrinsic column-density scatter in the data. Remarkably, the column-density residuals in the right panel now do not show any abnormal trend of $\delta$log$N$ increasing with $b$, indicating that our two-component model is likely a better representation of the \SiIV\ seen toward QSOs than the commonly adopted flat-slab model. Because our model does not have distance information from the QSO sample, we cannot yield meaningful constraints on the spatial extensions of the disk-halo and global components. We discuss the potential origin(s) of the global component in \S\ \ref{sec5.2}.

\section{Discussion}
\label{sec5}

In the previous section, we briefly reviewed the commonly adopted flat-slab model and proposed a two-component model to better describe the all-sky distribution of \SiIV\ absorption at low-intermediate velocity. Here,  we first discuss the caveats of both models (\S\ \ref{sec5.1}). We proceed by discussing the potential origin(s) of the \SiIV\ global component, which dominates over the disk-halo component in column density (\S\ \ref{sec5.2}). Finally, we conduct order-of-magnitude mass calculations for the global component in the context of the MW's CGM (\S\ \ref{sec5.3}).

\subsection{Caveats of the Models}
\label{sec5.1}

A considerable degree of column-density scatter can be seen in both \citetalias{Savage09}'s star/QSO sample and our QSO sample, indicating that the \SiIV-bearing gas is patchy. There are several reasons for the patchiness of a disk-halo and halo medium. For example, the Galactic fountain \citep{Shapiro76} that circulates material between the disk and halo may be responsible for some of the scatter.  Along the trajectory of a fountain gas parcel, the gas may experience several phase changes due to different cooling rates at different locations \citep{Marasco12}, resulting in a clumpy and inhomogeneous medium. Additional processes may also contribute to the column-density scatters, such as outflowing material from star-forming sites (e.g., \citealt{Rubin14}), inflowing filaments from the halo (e.g., \citealt{Zheng17a}) as well as the debris of satellites passing through the Galactic halo (e.g, the Magellanic Stream; \citealt{Nidever10, Fox14}).

The concept of a patchiness parameter was originally introduced by \citetalias{Savage09}. To obtain the best parameters for their flat-slab disk-halo model, \citetalias{Savage09} incorporated a fixed logarithmic $\sigma_p$ value to their measurement errors so that the reduced $\chi^2$ could be forced to 1.0 at their best fits. Without $\sigma_p$, the reduced $\chi^2$ would be $\gg$1.0 and the best-fit parameters would be determined by those measurements with the smallest errors.  In their setup, $\sigma_p$ is a mathematical remedy for the oversimplified model fitting; however, it does not convey a clear physical meaning. Our two-component model defines the patchiness parameter $\sigma_p$ as to reflect the intrinsic column-density scatter from sightline to sightline. The intrinsic scatter is assumed to follow a Gaussian distribution with a variance of $\sigma_p^2$. This setup of $\sigma_p$ captures the large fluctuations in the dataset, however, it still lacks a physically motivated meaning of the nature of gas structures over the Galactic sky.

Another caveat pertaining to both \citetalias{Savage09}'s and our model is rooted in the sample selection. The halo stars in \citetalias{Savage09} lie mostly at $|b|<15\degree$ and toward the inner Galaxy direction (see their figure 1). Their sightlines at higher Galactic latitudes consist of 15 QSOs at $b<45\degree$ and six stars in the LMC/SMC. Along the QSO and LMC/SMC stellar sightlines, the sources of and distances to the foreground low-intermediate velocity absorbers are not well-constrained. This inhomogeneous sample selection and the inclusion of distant QSOs and LMC/SMC stellar sightlines in their model fitting likely overestimate the flat-slab scale height. In contrast, our sample is dominated by QSOs at $|b|\gtrsim30\degree$ (see Figures \ref{fig4}/\ref{fig5}) and our two-component model does not have distance information on the gas as limited by the QSO data sample.


We note that our model finds a significant global background in addition to the disk-halo component, whereas \citetalias{Savage09}'s flat-slab model does not include such a component. The inconsistency between the two models suggests that the gaseous structures of the MW cannot be simply parameterized by either a flat-slab or a global distribution. \citetalias{Savage09}'s flat-slab model captures the properties of the MW's disk-halo interface at low Galactic latitudes, whereas our two-component model better represents the MW halo at higher latitudes. Ideally, the QSOs at high latitudes and stars near the Galactic plane should be combined to provide better modeling constraints on the distribution of the warm-ionized gas in the MW halo.



\subsection{Origin(s) of the \SiIV\ Global Component}
\label{sec5.2}

Here we discuss the origin(s) of the global component revealed in \S\ \ref{sec4.2}. We attempt to consider all of the possible ionized structures that the QSO paths may intercept.

First, we consider a Galactic ISM origin. It is unlikely that our QSO sightlines would suffer strong contamination from dense spiral-arm regions or warps at the outskirts of the MW's disk because the QSOs are mostly at $|b|\gtrsim30\degree$. However, all sightlines must leave the disk through the Local Bubble, a cavity in the ISM at the location of the Sun with a radius of $\sim100$ pc. \citetalias{Savage09}'s halo star sample contains three objects within 1 kpc of the Sun (HD\_38666, HD\_52463, HD\_116658). The mean \SiIV\ column density measured along these sightlines is log$N\sim$12.5 dex, which accounts for only $\sim$10\% of the \SiIV\ toward distant QSOs. Furthermore, \cite{Savage06} found that the local bubble contributes only $\sim9\%$ of the column density to the total \OVI\ content in the MW halo by comparing UV spectra of nearby white dwarfs at a mean distance of $\sim109$ pc with those of MW halo stars. On the basis of this result, \citetalias{Savage09} suggest that \SiIV\ from the Local Bubble would not have a significant effect, despite the potential phase difference between the \OVI- and \SiIV-traced material.

Second, superbubbles breaking out of the ISM near the solar neighborhood are unlikely to be the source. This is primarily because the upstreaming material should be found with mostly positive radial velocities, in stark contrast to our velocity-centroid measurements (Figure \ref{fig4}) that predominantly trace negative radial velocities. However, it is possible that we are seeing the remnants of a past superbubble break-out, where the bulk of the previously ejected material is now falling back down. In this scenario, the fall-back gas is not distinguishable from the observed ionized intermediate- and high-velocity clouds, which we will discuss shortly.

Third, as our sightlines cover most of the Galactic sky at $|b|\gtrsim30\degree$, any particular substructures covering small areas of the sky cannot be the dominant sources, but are likely to contribute to the column-density scatter. The potential structures include: (1) the north and south Fermi Bubbles (e.g., \citealt{Fox15,Bordoloi17}) which we have discussed in \S\ \ref{sec3}, (2) the CGM of M31 in the southern sky \citep{Lehner15}, and (3) cool filamentary structures embedded in the $10^5-10^7$ K warm-hot intergalactic medium in the Local Group (e.g., \citealt{Nicastro03, Richter17}).  As we divide the sky into octants for block bootstrapping, the influence of sightlines potentially associated with small-area structures have been reflected in the distribution of the ($N_{\rm DH, \perp}$, $N_{\rm G}$) solutions (Figure \ref{fig9}).


Fourth, we consider that the global component may be associated with the widespread \HI\ structures observed in the inner Galactic halo. Intermediate-velocity clouds are defined as discrete \HI\ clouds moving at $|v|<100$ \kms\ \citep{vanWoerden04, Albert04, Wakker04} at distances of $\sim1-5$ kpc above the Galactic plane (see Table 2 in \citealt{Albert04}), and high-velocity clouds are those moving at $|v|>100$ \kms\ at further distances around $\sim10$ kpc \citep{vanWoerden04,Putman12, Lehner12, Richter17}. Because we only consider the origin of the \SiIV\ global component at low-intermediate velocities, ionized HVCs themselves are not the possible culprits; however, clouds have been found with properties similar to HVCs moving at lower velocities \citep{Peek09, Saul14} as expected from models \citep{Wakker91}. If the global component represents the ionized counterparts of the IVCs or HVCs, one would expect the \SiIV-bearing clouds to be distributed in a spherical configuration so that the column densities have minimal dependence on $|b|$.


Ultimately, we consider that the global component may represent a contribution from material occupying the MW's CGM, extending to $\sim0.5$ $R_{\rm vir}$. All QSO sightlines would intercept the extended CGM material of the MW. As mentioned in the Introduction, the MW's CGM is typically studied at high velocities ($|v|>$100 \kms) to avoid contamination from more nearby low-intermediate velocity gas (e.g., \citealt{Fox06,Shull09,Collins09,Lehner12,Richter17}). However, \cite{Zheng15} found that $\sim65\%$ of the MW's CGM mass above $|b|\geq20\degree$ is likely to be hidden at $|v|\leq100$ \kms\ regardless of gas phases, according to a synthetic observation of a simulated MW-mass galaxy from a cosmological simulation \citep{Joung12}. Empirically, as found in extragalactic CGM studies, highly ionized, metal-enriched gas can extend to nearly half the virial radii of the host galaxies with centroid velocities largely within $|v|\lesssim200$ \kms\ of the systemic velocities (e.g., \citealt{Werk14, Liang14,Burchett16, Keeney17}). Therefore, it is likely that the \SiIV\ in the MW's CGM extends beyond the inner Galactic halo and casts absorption features in the low-intermediate velocity range.


To summarize, we find that the global component is unlikely to be dominated by the Local Bubble, superbubble breakouts, Fermi Bubble, M31's CGM, or the Local Group medium. The global component is most likely to represent the ionized counterparts to nearby IVCs/HVCs, and/or the MW's extended, ionized CGM that occupies a much larger volume, although its exact location is unknown. In each of these potential scenarios, spherical-like geometry better describes the data than the flat-slab cylindrical geometry because there is not an observed dependence on $|b|$ for the \SiIV\ column densities (Figures \ref{fig5}, \ref{fig7}, and \ref{fig10}). The dominant global component over the disk-halo component shows that there is likely to be a large amount of ionized gas moving at low-intermediate velocities hidden in the MW's CGM.

\subsection{Order-of-magnitude Calculation of \\the MW's CGM Mass}
\label{sec5.3}

Although our data show a covering fraction of 100\%\ over the Galactic sky, the distance to the gas in the extended halo is unclear. As found in extragalactic work \citep{Werk13, Keeney17}, the radial profile of gas density in the CGM drops as a power law: the gas in the inner halo is denser than the gas in the outskirts. \cite{Werk13} suggested that for star-forming galaxies with specific star formation rate $\geq10^{-11}$ yr$^{-1}$, the covering fraction of \SiIV\ is 94\%\ within an impact parameter of $\leq75$ kpc but drops to 58\%\ for regions between 75 and 160 kpc. This indicates that for observations of gas in the MW's halo, the column density along the LOS is most likely to be dominated by nearby gas in the inner Galactic halo.

Here we conduct an order-of-magnitude estimate for the mass of the low-intermediate velocity \SiIV-bearing gas in the MW's CGM. We consider the case that the global component found in \S\ \ref{sec4.2} is associated with the MW’s extended CGM in a spherical volume with a constant-density profile $n(r)\equiv N_{\rm G}/R$, where $R$ is the size of the halo. The total mass associated with this component is $M_{\rm SiIV} = \frac{4\pi}{3} R^3 (\frac{N_{\rm G}}{R}) m_{\rm SiIV}C_f$, where $N_{\rm G}=3.4\times10^{13}$ cm$^{-2}$ is the median column-density value of the global component we derived in \S\ \ref{sec4.2.2}, $m_{\rm SiIV}$ the mass of a \SiIV\ atom, and $C_f=100\%$ the covering fraction. For the radius, we adopt $R=75$ kpc on the basis of the covering fraction study of \cite{Werk13}, as mentioned above. The choice of $R=75$ kpc results in a conservative mass estimate for the \SiIV-bearing gas in the MW's CGM since there could be more diffuse gas beyond this volume but within the virial radius. The mass of the low-intermediate velocity \SiIV\ in the MW's CGM is:
\begin{equation}
M_{\rm SiIV} \gtrsim 1.8 \times 10^5 M_\odot (\frac{R}{75\ {\rm kpc}})^2 (\frac{N_{\rm G}}{3.4\times10^{13}\ {\rm cm}^{-2}})( \frac{C_f}{1}).
\label{eq12}
\end{equation}
We further convert the \SiIV\ mass into a total gas mass by assuming that the \SiIV-bearing gas represents the bulk property of the MW's CGM at cool phases:
\begin{equation}
\begin{array}{l l}
M_{\rm gas}&=1.4 m_{\rm H} (\frac{M_{\rm SiIV}}{m_{\rm Si}}) f_{\rm SiIV}^{-1} {\rm (Si/H)_{\odot}^{-1} Z^{-1}}\\
              &\gtrsim 3.1 \times 10^{9}\ M_{\odot} (\frac{f_{\rm SiIV}}{0.3})^{-1}(\frac{C_f}{1})(\frac{R}{75\ {\rm kpc}})^2\\
              &\ \ \ (\frac{\rm Z}{\rm 0.3\ Z_{\odot}})^{-1} (\frac{N_{\rm G}}{3.4\times10^{13}\ {\rm cm}^{-2}}),
\end{array}
\label{eq13}
\end{equation}
where 1.4 is to account for the presence of helium and $m_{\rm H}$ is the mass of a hydrogen atom. We adopt ${\rm (Si/H)_{\odot}}=10^{-4.49}$ from \cite{Asplund09}, which is the silicon abundance in the present-day solar photosphere relative to hydrogen. We assume the metallicity as Z $=$ 0.3 Z$_{\odot}$, on the basis of the COS-Halos mean CGM metallicity estimate \citep{Prochaska17}. The most uncertain term is the ionization fraction $f_{\rm SiIV}$, defined as the ratio of \SiIV\ to all the silicon in various neutral and ionized states. Photoionization, collisional ionization, and other non-equilibrium processes such as  shock ionization, conductive interfaces, and turbulence mixing may contribute to producing \SiIV\ (e.g., \citealt{Fox05}; \citetalias{Wakker12}). We examine the ionization fraction produced by CLOUDY (v13.03; \citealt{Ferland98, Ferland13}) in an extragalactic UV background \citep{Haardt01} at \HI\ column density of $10^{20}$ cm$^{-2}$, and find $f_{\rm SiIV}$ peaks at $\sim0.3$ for Z $=0.1-0.3$ Z$_{\odot}$. This also results in a conservative mass estimate becauses we assume a maximum ionization fraction of \SiIV. Because we only aim for an order-of-magnitude mass calculation, we do not conduct sophisticated ionization modelings to determine $f_{\rm SiIV}$. We caution that the \SiIV\ ionization fraction is highly uncertain without fully understanding its ionization mechanism (e.g. \citetalias{Wakker12}; \citealt{Werk16}). Furthermore, metallicity measurements for the ionized gas are notoriously difficult to obtain. Large systematic uncertainties in gas ionization state, metallicity, and medium clumpiness plague all mass estimates based on ionized gas absorption lines. Therefore, our mass calculation for the MW’s extended CGM is accurate only as an order-of-magnitude estimate.

\cite{Zheng15} suggested that the mass ratio of high- to low-intermediate velocity gas in the MW halo at $|b|\geq 20\degree$ is $0.35:0.65$ and this mass ratio does not vary significantly from 10 kpc to 250 kpc, on the basis of a synthetic observation of a simulated MW-mass galaxy. If we apply this ratio to the mass in Eqn. \ref{eq12} and \ref{eq13}, we find the total \SiIV\ and hydrogen mass in the MW's CGM is,
\begin{equation}
\begin{array}{l l}
M_{\rm SiIV, all}&\gtrsim M_{\rm SiIV}/0.65 \gtrsim 2.8\times10^5\ M_{\odot} \\
M_{\rm gas, all}&\gtrsim M_{\rm gas}/0.65 \gtrsim 4.7\times10^9\ M_{\odot}.
\end{array}
\end{equation}

In addition, we can infer the mass for the high-velocity gas ($|v_{\rm LSR}|> 100$ km s$^{-1}$) at cool phases in the Galactic halo, which will be $M_{\rm gas, hvc}=0.35M_{\rm gas, all}\gtrsim 1.7\times10^9\ M_{\odot}$. When compared to the literature, We find that $M_{\rm gas, hvc}$ is consistent with the mass estimate for the MW's ionized HVCs including the Magellanic Stream ($M\geq3.0\times10^9 M_{\odot}$; \citealt{Fox14, Richter17}). We also compare our result with extragalactic CGM studies and find that our derived mass of the MW's CGM ($M_{\rm gas, all}$) is between the estimate from the COS-Halos team \citep{Werk14, Prochaska17} and that of M31's CGM by \cite{Lehner17}. \cite{Prochaska17} found the total mass of the cool, photoionized CGM of $L\sim L*$ galaxies at redshift $\sim0.2$ is $(9.2\pm4.3)\times10^{10}\ M_{\odot}(\frac{R}{160\ \rm kpc})^2$, which is $(2.0\pm0.9)\times10^{10}\ M_{\odot}(\frac{R}{75\ \rm kpc})^2$ for our assumed volume. On the other hand, \cite{Lehner17} derived a total gas mass of $2\times10^8 M_{\odot}(\frac{R}{50 {\rm kpc}})^2(\frac{Z}{Z_\odot})^{-1}$ for M31's CGM by considering \SiII, \SiIII, and \SiIV\ simultaneously within 50 kpc of the galaxy. This corresponds to a total gas mass of $\sim1.5\times10^9 M_\odot (\frac{R}{75\ {\rm kpc}})^2(\frac{Z}{0.3\ {\rm Z_{\odot}}})^{-1}$ adopting our assumed volume and metallicity for the MW's CGM.

We note that other methods, such as the shell-geometry and volume-filling factor, are also used to calculate CGM masses. For example, \cite{Werk14} computed the CGM mass for COS-Halos galaxies by integrating the gas surface-density profile in an annular manner (see also \citealt{Tumlinson11, Bordoloi14, Stern16, Prochaska17}; etc.). If we apply this method and assume the \SiIV-bearing gas is distributed over a spherical shell with radius of 75 kpc, the mass estimate for the cool gas in the MW CGM will be a factor of three higher. On the other hand, \cite{Stocke13} estimated the CGM mass of late-type galaxies with a volume filling factor of 3\%$-$5\%. Briefly, they determined the volume-filling factor of warm CGM clouds by taking into account the covering fraction, the average number of ion absorption components along the sightlines, and the physical cloud sizes as inferred from photoionization modelings (see their equation 5). The total CGM mass was derived by summing up the mass of all available clouds with the assumption that warm clouds represent the whole CGM cloud population. \cite{Werk14} calculated the CGM mass for the COS-Halos sample with both the surface-density method (as mentioned above) and the volume-filling method. With a volume-filling factor of 11\%, they found consistent CGM mass results between the two methods. The volume-filling method has two requirements: (1) an estimate of cloud sizes based on CLOUDY modeling, and (2) an estimate of the average number of ion absorption components along the sightlines based on Voigt-profile fitting. Because neither of these is used in this work, we are unable to conduct mass estimate with the volume-filling method.

\section{Conclusion}
\label{sec6}

For MW absorption-line studies that rely on background QSOs, the LOSs unavoidably intercept the ISM, the disk-halo interface, and the CGM of the MW. Therefore, every QSO spectrum contains mixed absorption-line features from these sources. High-velocity gas ($|v_{\rm LSR}|>100$ \kms) is traditionally assumed to relate to the CGM, whereas low-intermediate velocity gas ($|v_{\rm LSR}|\leq100$ \kms) is considered to be more nearby. In this work, we study the all-sky distribution and origins of the \SiIV-bearing gas at low-intermediate velocity with coadded G130M COS spectra of 132 QSOs obtained from HSLA DR1 \citep{Peeples17}. Our results are summarized as follows.

First, the low-intermediate velocity \SiIV\ measured toward all-sky QSOs has an average column density of $\langle N\rangle = (3.8\pm1.4)\times10^{13}$ cm$^{-2}$ (i.e., log$N=13.58\pm1.16$ dex). The \SiIV\ column density does not significantly correlate with the Galactic latitude or longitude (see \S\ \ref{sec3}). We find that most sightlines in the north show gas moving at negative velocities toward the LSR at $-50\lesssim v \lesssim0$ \kms, which implies potential inflow. Sightlines in the south do not show such velocity preference.

Second, we examine our QSO data using a commonly adopted plane-parallel flat-slab model, but find that the flat-slab model cannot explain the log$N$ distribution as a function of $b$ (see \S\ \ref{sec4.1}). The flat slab predicts higher column densities at lower Galactic latitudes because the LOS paths going through the slab are longer; however, as summarized in the previous point, our data do not show such $b$-dependent trend.

Third, we propose a two-component model that consists of a disk-halo and a global component (see \S\ \ref{sec4.2}). The disk-halo component $\NDH$ is assumed to follow the flat-slab geometry; thus the LOS column-density scales with 1/sin$|b|$. The global component $N_{\rm G}$ represents a general background that contributes identically from sightline to sightline. Additionally, we consider a patchiness parameter drawn from a Gaussian distribution (0, $\sigma_p^2$) to account for the intrinsic column density from sightline to sightline. We conduct Bayesian analyses and block bootstrapping to solve for ($\NDH$, $N_{\rm G}$, $\sigma_p$), and find $\NDH=1.3^{+4.7}_{-0.7}\times10^{12}$ cm$^{-2}$, $N_{\rm G}=3.4^{+0.3}_{-0.3}\times10^{13}$ cm$^{-2}$, and $\sigma_p=1.4\times10^{13}$ cm$^{-2}$, which are consistent with the statistics measured from the data. Our two-component model is able to reproduce the column-density distribution seen toward distant QSOs (Figure \ref{fig10}).

Ultimately, we examine the origin(s) of the global component (\S\ \ref{sec5.2}) and find that it is most likely to be associated with the MW's CGM, although its exact location is difficult to constrain without distance measurements. If it were related to the MW's CGM, we would find a total mass of $M_{\rm gas, all}\gtrsim4.7\times10^9\ M_{\odot} (\frac{C_f}{1})(\frac{R}{75\ {\rm kpc}})^2 (\frac{f_{\rm SiIV}}{0.3})^{-1}(\frac{Z}{0.3\ Z_{\odot}})^{-1}$ for gas at cool phases, including the mass of low-intermediate velocity gas ($|v_{\rm LSR}|\leq100$ km s$^{-1}$) directly probed in this work and that of the high-velocity gas ($|v_{\rm LSR}|>100$ km s$^{-1}$) inferred from a synthetic observation of a simulated MW-mass galaxy. Our mass estimates are consistent with MW ionized HVC studies and  extragalactic CGM studies. This work shows that there is likely to be a considerable amount of ionized gas moving at low-intermediate velocity hidden in the MW's CGM.

In conducting this study, we perform continuum normalization for a total of 401 QSO spectra observed with HST/COS  (\S\ \ref{sec2.1}), which is based on a QSO subset of HSLA DR1 \citep{Peeples17}. Each coadded spectrum has G130M and/or G160M gratings and has a S/N $\geq$ 5.0 per resolution element over the wavelength span. Our continuum normalization focuses on Galactic interstellar absorption lines, including \FeII\ 1142/1143/1144/1608 \AA, \PII\ 1152 \AA, \SII\ 1250/1253/1259 \AA, \SiII\ 1190/1193/1260/1526 \AA, \SiIII\ 1206 \AA, \SiIV\ 1393/1402 \AA, \CII\ 1334 \AA, and \CIV\ 1548/1550 \AA. For each QSO we also retrieve \HI\ 21cm emission lines from three \HI\ surveys: GALFA-\HI\ \citep{Peek18}, LAB \citep{Kalberla05}, and HI4PI \citep{HI4PI}. The QSO continua, individual line spectra, and the corresponding \HI\ 21cm lines together form our COS-GAL dataset, which we release to the public at \href{http://dx.doi.org/10.17909/T9N677}{[10.17909/T9N677]}. In addition, the codes and data used in this work can be found at \href{https://github.com/yzhenggit/Zheng18_MWCGM}{https://github.com/yzhenggit/Zheng18\_MWCGM}.

Critical information is missing when relying on QSOs to probe the MW halo -- the distance. As we mention in \S\ \ref{sec5.1}, ideally the stellar sample and the QSO sample should be combined to yield better constraints on the location to the \SiIV\ content associated with the MW's disk-halo interface, thus shedding light on the distribution of the low-intermediate velocity gas hidden in the MW's CGM. The COS-GAL dataset can be further used to conduct preliminary studies of other interstellar lines related to the MW's ISM and halo components, such as studying the global phenomena of ionized HVCs, gas flows in different directions, the ionized gas properties in specific regions of the MW, or the surroundings of galaxies in the Local Group.

{\it Acknowledgement.} We thank the anonymous referees for their patient and useful comments in improving this manuscript. We are very grateful to B. Wakker and B. Savage for pointing out a critical systemic error in calculating the \SiIV\ apparent column density in the first version. We thank K. Tchernyshyov, M. Peeples, S. Flemming, H.W. Chen, A. Fox, and Z.J. Qu for useful discussions on many parts of this manuscript. Y.Z. is partially supported by the Miller Institute for Basic Research in Science as of finishing this manuscript. JW acknowledges support from a 2018 Alfred P. Sloan Research Fellowship. We acknowledge support from HST-GO-13706, HST-GO-13383, and HST-GO-13382 which were provided by NASA through grants from the Space Telescope Science Institute (STScI). STScI is operated by the Association of Universities for Research in Astronomy, Inc., under NASA contract NAS5-26555. Most of the data presented in this paper were obtained from the Mikulski Archive for Space Telescopes (MAST). We also acknowledge support from the National Science Foundation under Grant No. AST-1410800 and AST-1312888.

\software{Astropy \citep{Astropy13}, Linetools \citep{linetools}, IPython package \citep{Ipython07}, matplotlib \citep{Hunter07}, Emcee \citep{emcee}, Github.}

\bibliographystyle{apj}
\bibliography{main}

\begin{turnpage}
\begin{table}
\renewcommand{\arraystretch}{0.9}
\caption{\SiIV\ Data Sample}
\begin{center}
\begin{tabular}{llrrrrrrrrrl}
\hline
\hline
No.&                   HSLA-ID&      Glon    &     Glat   &        $z$&   S/N& $Q$ &  log$N_{1393}$  &  $v_{c, 1393}$  &  log$N_{1402}$  &  $v_{c, 1402}$  & Simbad ID\\
   &                          &($\degree$) &($\degree$)&         &  & &                 &  (km s$^{-1}$)    &                 &  (km s$^{-1}$)    &          \\
(0)&                   (1)    &      (2)   &     (3)   &   (4)   &   (5)  &(6)&            (7)  &  (8)            &         (9)     &  (10)           & (11)     \\
\hline
  1&                   NGC-5548&     31.9607&     70.4957&   0.0163&  154.2&   0&  13.75$\pm$ 0.00&  -43.8$\pm$  0.3&  13.75$\pm$ 0.00&  -45.2$\pm$  0.5&  NGC 5548 \\
  2&                    MARK509&     35.9711&    -29.8553&   0.0341&   92.8&   0&  13.83$\pm$ 0.00&   22.6$\pm$  0.3&  13.83$\pm$ 0.00&   26.6$\pm$  0.6&  Mrk 509 \\
  3&                     MRK876&     98.2694&     40.3758&   0.1385&   84.9&   0&  13.64$\pm$ 0.00&  -24.6$\pm$  0.5&  13.64$\pm$ 0.01&  -20.5$\pm$  1.0&  Mrk 876 \\
  4&             RXJ1230.8+0115&    291.2605&     63.6602&   0.1170&   61.6&   0&  13.53$\pm$ 0.01&   -5.8$\pm$  0.8&  13.54$\pm$ 0.01&   -4.3$\pm$  1.6&  2MASS J12305003+0115226 \\
  5&                 PG0804+761&    138.2787&     31.0327&   0.1000&   59.7&   0&  13.58$\pm$ 0.01&  -37.3$\pm$  0.9&  13.60$\pm$ 0.01&  -35.7$\pm$  1.8&  2MAXI J0808+757 \\
  6&                   NGC-7469&     83.0985&    -45.4670&   0.0159&   58.4&   0&  13.50$\pm$ 0.01&   -0.8$\pm$  0.9&  13.50$\pm$ 0.01&   -2.6$\pm$  1.7&  NGC 7469 \\
  7&                 PG1116+215&    223.3600&     68.2094&   0.1765&   47.4&   0&  13.68$\pm$ 0.01&  -12.9$\pm$  0.9&  13.69$\pm$ 0.01&  -11.2$\pm$  1.4&  Ton 1388 \\
  8&           IRAS-F22456-5125&    338.5115&    -56.6293&   0.1000&   47.3&   0&  13.33$\pm$ 0.01&    6.9$\pm$  1.6&  13.34$\pm$ 0.03&    8.4$\pm$  3.2&  2MASS J22484115-5109532 \\
  9&                PKS2155-304&     17.7305&    -52.2458&   0.1160&   45.7&   0&  13.38$\pm$ 0.01&    0.3$\pm$  1.5&  13.36$\pm$ 0.02&   -1.6$\pm$  3.1&  QSO B2155-304 \\
 10&                 PG1352+183&      4.3748&     72.8738&   0.1515&   44.5&   0&  13.65$\pm$ 0.01&  -34.2$\pm$  1.3&  13.64$\pm$ 0.02&  -39.5$\pm$  2.4&  2E 1352.2+1820 \\
 11&                     IO-AND&    122.2786&    -23.1812&   0.1340&   43.0&   0&  13.26$\pm$ 0.02&  -28.5$\pm$  2.2&  13.26$\pm$ 0.03&  -10.0$\pm$  4.1&  2MASX J00481899+3941118 \\
 12&                     MRK817&    100.2997&     53.4783&   0.0315&   42.9&   0&  13.47$\pm$ 0.01&  -33.2$\pm$  1.6&  13.48$\pm$ 0.02&  -30.0$\pm$  2.9&  Mrk 817 \\
 13&           IRAS-L06229-6434&    274.3116&    -27.3194&   0.1290&   42.4&   0&  13.35$\pm$ 0.01&   12.7$\pm$  1.9&  13.33$\pm$ 0.03&   23.6$\pm$  4.1&  [VV98] J062309.1-643624 \\
 14&                    TONS210&    224.9716&    -83.1603&   0.1160&   42.0&   0&  13.63$\pm$ 0.01&  -15.7$\pm$  1.0&  13.64$\pm$ 0.02&  -14.5$\pm$  2.0&  Ton S 210 \\
 15&                    NGC-985&    180.8371&    -59.4903&   0.0427&   40.6&   0&  13.35$\pm$ 0.01&  -39.1$\pm$  2.2&  13.36$\pm$ 0.03&  -31.4$\pm$  4.0&  NGC 985 \\
 16&                 PG0052+251&    123.9075&    -37.4377&   0.1550&   40.3&   0&  13.35$\pm$ 0.02&  -17.6$\pm$  2.1&  13.37$\pm$ 0.02&  -19.0$\pm$  2.9&  2E 217 \\
 17&                    PHL1811&     47.4735&    -44.8151&   0.1940&   39.6&   0&  13.61$\pm$ 0.01&    1.2$\pm$  1.0&  13.60$\pm$ 0.02&    2.0$\pm$  2.0&  QSO J2155-0922 \\
 18&                PG-1407+265&     34.6686&     72.5886&   0.9400&   36.6&   0&  13.78$\pm$ 0.01&  -43.1$\pm$  1.3&  13.77$\pm$ 0.01&  -39.1$\pm$  2.1&  QSO J1409+2618 \\
 19&                HE0226-4110&    253.9406&    -65.7747&   0.4934&   36.0&   0&  13.36$\pm$ 0.02&   -0.8$\pm$  2.0&  13.36$\pm$ 0.03&    8.3$\pm$  4.0&  HE 0226-4110 \\
 20&                      3C263&    134.1590&     49.7439&   0.6520&   35.8&   0&  13.70$\pm$ 0.01&  -44.7$\pm$  1.5&  13.71$\pm$ 0.02&  -39.9$\pm$  2.4&  7C 113710.40+660425.00 \\
 21&                   FAIRALL9&    295.0729&    -57.8265&   0.0480&   35.6&   0&  13.59$\pm$ 0.01&   12.9$\pm$  1.3&  13.60$\pm$ 0.02&   20.9$\pm$  2.6&  ESO 113-45 \\
 22&                 PG0953+414&    179.7860&     51.7091&   0.2341&   35.4&   0&  13.38$\pm$ 0.02&  -21.8$\pm$  2.1&  13.41$\pm$ 0.03&  -13.2$\pm$  3.8&  2XMM J095652.4+411522 \\
 23&              QSO-B1307+085&    316.7864&     70.7068&   0.1543&   35.2&   0&  13.73$\pm$ 0.01&  -21.5$\pm$  1.2&  13.71$\pm$ 0.01&  -21.2$\pm$  1.6&  2E 2978 \\
 24&                 PG1259+593&    120.5554&     58.0479&   0.4778&   34.7&   0&  13.36$\pm$ 0.02&  -18.1$\pm$  2.2&  13.37$\pm$ 0.03&  -10.6$\pm$  4.3&  LB 2522 \\
 25&                HE0056-3622&    293.7190&    -80.8980&   0.1641&   34.4&   0&  13.63$\pm$ 0.01&   -9.4$\pm$  1.3&  13.64$\pm$ 0.02&   -3.0$\pm$  2.0&  6dFGS gJ005837.4-360605 \\
 26&                   VIIZW244&    136.6575&     32.6778&   0.1313&   34.1&   0&  13.59$\pm$ 0.01&  -45.7$\pm$  1.9&  13.61$\pm$ 0.02&  -35.4$\pm$  2.9&  MCG+13-07-002 \\
 27&                1ES1553+113&     21.9089&     43.9642&   0.3600&   34.0&   0&  13.96$\pm$ 0.01&   -3.5$\pm$  0.6&  13.95$\pm$ 0.01&   -7.6$\pm$  1.1&  QSO B1553+113 \\
 28&                 ESO-141-55&    338.1834&    -26.7112&   0.0370&   33.4&   0&  13.80$\pm$ 0.01&    4.2$\pm$  0.9&  13.81$\pm$ 0.01&    4.2$\pm$  1.6&  ESO 141-55 \\
 29&                 S50716+714&    143.9811&     28.0176&   0.3000&   33.4&   0&  13.47$\pm$ 0.01&  -34.8$\pm$  2.1&  13.45$\pm$ 0.03&  -42.3$\pm$  4.5&  7C 071610.69+712601.00 \\
 30&                 PG1626+554&     84.5149&     42.1886&   0.1330&   33.2&   0&  13.53$\pm$ 0.01&  -24.1$\pm$  1.7&  13.55$\pm$ 0.02&  -24.1$\pm$  3.1&  QSO B1626+5529 \\
 \hline
 \hline
 \end{tabular}
 \end{center}
 \vspace{-2.5mm}
 \tablenotetext{}{Note - Col (0): QSOs are arranged from high to low S/N per resolution element. Col (1): QSO ID as adopted by HSLA first data release \citep{Peeples17}. Col (2) and (3): Galactic longitude and latitude as provided on Simbad. 
 Col (4): QSO redshift as provided on Simbad. Col (5): S/N per resolution element for the absorption-free region between 1394\AA\ and 1401\AA. The S/N value can be calculated as S/N per pix$\times N/\sqrt{N}$, where $N$ is the total number of pixels for a corresponding resolution. For G130M grating, $N=6$. Col (6): Quality flagging (see \S\ \ref{sec2.2}): $Q=0$ means this target has spectrally resolved doublet profiles, $Q=1$ means its \SiIV\ lines are saturated, and $Q=-1$ means its 1393 \AA\ line is abnormally stronger than 1402 \AA. Col (7) and (8): logarithmic apparent column density and centroid velocity for \SiIV\ 1393\AA\ integrated from -100 to 100 \kms. Col (9) and (10): Same as Col (7) and (8), but for \SiIV\ 1402 \AA. Col (11): Simbad ID. The machine readable version of this table can be downloaded here: \href{https://github.com/yzhenggit/Zheng18_MWCGM/}{https://github.com/yzhenggit/Zheng18\_MWCGM/}.}
 \label{tb1}
 \end{table}
 \end{turnpage}

\begin{turnpage}
\begin{table}
\renewcommand{\arraystretch}{0.9}
\caption*{Table 1 continued}
\begin{center}
\begin{tabular}{llrrrrrrrrrl}
\hline
\hline
No.&                   HSLA-ID&      Glon    &     Glat   &        $z$&   S/N& $Q$ &  log$N_{1393}$  &  $v_{c, 1393}$  &  log$N_{1402}$  &  $v_{c, 1402}$  & Simbad ID\\
   &                          &($\degree$) &($\degree$)&         & & &                 &  (km s$^{-1}$)    &                 &  (km s$^{-1}$)    &          \\
(0)&                   (1)    &      (2)   &     (3)   &   (4)   &   (5)  &(6)&            (7)  &  (8)            &         (9)     &  (10)           & (11)     \\
\hline
 31&                 MR2251-178&     46.1973&    -61.3256&   0.0640&   32.8&   0&  13.65$\pm$ 0.01&    9.9$\pm$  1.1&  13.65$\pm$ 0.02&   13.5$\pm$  2.4&  NAME MR 2251-178 \\
 32&    SDSSJ135341.03+361948.0&     71.6687&     73.9144&   0.1470&   32.2&   0&  13.36$\pm$ 0.02&  -39.1$\pm$  2.5&  13.37$\pm$ 0.04&  -44.6$\pm$  6.9&  2MASS J13534104+3619480 \\
 33&             LBQS-1435-0134&    348.7185&     51.3746&   1.3077&   31.1&   0&  13.75$\pm$ 0.01&  -21.4$\pm$  1.2&  13.76$\pm$ 0.02&  -19.9$\pm$  2.0&  LBQS 1435-0134 \\
 34&            IRASF00040+4325&    114.4161&    -18.4214&   0.1660&   31.0&   0&  13.20$\pm$ 0.03&  -29.5$\pm$  4.0&  13.21$\pm$ 0.04&  -22.5$\pm$  6.3&  IRAS 00040+4325 \\
 35&                    MRK-335&    108.7635&    -41.4244&   0.0250&   31.0&   0&  13.59$\pm$ 0.01&   -9.1$\pm$  1.4&  13.61$\pm$ 0.02&   -6.6$\pm$  2.6&  Mrk 335 \\
 36&                   NGC-3783&    287.4560&     22.9474&   0.0098&   30.8&   0&  13.75$\pm$ 0.01&   11.1$\pm$  1.1&  13.76$\pm$ 0.01&    9.8$\pm$  1.8&  NGC 3783 \\
 37&                1H-2129-624&    331.1427&    -42.5233&   0.0590&   30.3&   0&  13.65$\pm$ 0.01&   29.0$\pm$  1.5&  13.66$\pm$ 0.02&   29.7$\pm$  2.8&  2MAXI J2136-624 \\
 38&                     MRK106&    161.1389&     42.8794&   0.1228&   30.2&   0&  13.31$\pm$ 0.02&  -21.5$\pm$  2.8&  13.33$\pm$ 0.04&  -20.6$\pm$  6.0&  Mrk 106 \\
 39&            RXS-J23218-7026&    313.2923&    -44.8373&   0.3000&   29.4&   0&  13.63$\pm$ 0.01&   19.2$\pm$  1.5&  13.65$\pm$ 0.02&   26.8$\pm$  2.9&  2MASS J23215113-7026441 \\
 40&                PKS0552-640&    273.4656&    -30.6114&   0.6840&   28.8&   0&  13.42$\pm$ 0.02&   20.7$\pm$  2.4&  13.46$\pm$ 0.03&   31.8$\pm$  4.6&  2MASS J05522451-6402108 \\
 41&             RXJ2154.1-4414&    355.1794&    -50.8647&   0.3440&   28.8&   0&  13.58$\pm$ 0.01&   22.2$\pm$  1.5&  13.60$\pm$ 0.03&   28.5$\pm$  3.7&  2MASS J21545109-4414057 \\
 42&                HE0238-1904&    200.4807&    -63.6334&   0.6280&   27.0&   0&  13.47$\pm$ 0.02&  -11.7$\pm$  2.2&  13.50$\pm$ 0.03&    0.0$\pm$  3.8&  QSO B0238-1904 \\
 43&                    MRK1513&     63.6697&    -29.0695&   0.0620&   26.8&   0&  13.73$\pm$ 0.01&    2.9$\pm$  1.2&  13.74$\pm$ 0.02&    9.6$\pm$  2.3&  Mrk 1513 \\
 44&                 B0117-2837&    225.7305&    -83.6507&   0.3490&   26.4&   0&  13.63$\pm$ 0.01&   -7.0$\pm$  1.5&  13.64$\pm$ 0.03&   -6.7$\pm$  3.5&  2E 357 \\
 45&              QSO-B0026+129&    114.6377&    -49.2459&   0.1420&   26.4&   0&  13.41$\pm$ 0.01&  -16.1$\pm$  1.9&  13.44$\pm$ 0.04&   -6.8$\pm$  5.5&  2E 93 \\
 46&                 PG1126-041&    267.6305&     52.7454&   0.0600&   26.1&   0&  13.43$\pm$ 0.02&   11.2$\pm$  2.4&  13.46$\pm$ 0.03&   20.3$\pm$  4.6&  Mrk 1298 \\
 47&                 PG1011-040&    246.5010&     40.7487&   0.0580&   25.9&   0&  13.37$\pm$ 0.02&   11.2$\pm$  2.8&  13.41$\pm$ 0.04&   13.9$\pm$  5.1&  QSO B1011-040 \\
 48&                HE0153-4520&    271.7954&    -67.9751&   0.4510&   25.8&   0&  13.42$\pm$ 0.02&    4.1$\pm$  2.4&  13.45$\pm$ 0.04&   -7.0$\pm$  4.5&  [VV2000] J015513.2-450612 \\
 49&      2XMM-J141348.3+440014&     83.8280&     66.3534&   0.0870&   25.5&   0&  13.44$\pm$ 0.02&  -45.5$\pm$  3.3&  13.44$\pm$ 0.04&  -29.6$\pm$  5.7&  2MASS J14134834+4400141 \\
 50&                HE2347-4342&    336.0339&    -69.5737&   2.8850&   25.0&   0&  13.54$\pm$ 0.02&  -10.9$\pm$  2.2&  13.56$\pm$ 0.03&   -5.1$\pm$  4.0&  QSO B2347-4342 \\
 51&             RXSJ09565-0452&    243.3318&     37.0043&   0.1570&   24.8&   0&  13.67$\pm$ 0.01&   13.1$\pm$  1.8&  13.67$\pm$ 0.02&   14.6$\pm$  2.6&  2MASX J09563012-0453174 \\
 52&                     MRK304&     75.9893&    -34.2225&   0.0663&   24.8&   0&  13.82$\pm$ 0.01&   -9.0$\pm$  1.1&  13.83$\pm$ 0.02&   -7.0$\pm$  2.1&  Mrk 304 \\
 53&                 PG1048+342&    190.6025&     63.4374&   0.1671&   24.6&   0&  13.70$\pm$ 0.01&  -24.9$\pm$  1.8&  13.70$\pm$ 0.02&  -19.6$\pm$  2.6&  QSO B1048+342 \\
 54&             RXSJ00537+2232&    123.6380&    -40.3290&   0.1464&   24.6&   0&  13.38$\pm$ 0.02&  -18.9$\pm$  2.8&  13.41$\pm$ 0.06&  -22.0$\pm$  7.9&  2MASX J00534617+2232222 \\
 55&                     3C-66A&    140.1429&    -16.7669&   0.3400&   24.5&   0&  13.31$\pm$ 0.03&  -21.9$\pm$  3.6&  13.31$\pm$ 0.05&  -38.2$\pm$  7.9&  7C 021929.69+424830.00 \\
 56&                   NGC-4051&    148.8832&     70.0852&   0.0022&   24.5&   0&  13.67$\pm$ 0.02&  -20.6$\pm$  2.0&  13.65$\pm$ 0.02&  -22.7$\pm$  3.3&  NGC 4051 \\
 57&                       3C57&    173.0775&    -67.2618&   0.6710&   24.4&   0&  13.39$\pm$ 0.02&  -17.5$\pm$  2.9&  13.41$\pm$ 0.04&  -21.7$\pm$  5.7&  2MASS J02015718-1132334 \\
 58&                 PG0003+158&    107.3184&    -45.3258&   0.4505&   24.4&   0&  13.51$\pm$ 0.02&    7.4$\pm$  2.2&  13.53$\pm$ 0.03&   -2.6$\pm$  4.1&  2E 12 \\
 59&                    MRK1392&      2.7540&     50.2640&   0.0359&   24.2&   0&  13.71$\pm$ 0.01&  -22.7$\pm$  1.6&  13.73$\pm$ 0.02&  -24.1$\pm$  2.8&  Mrk 1392 \\
 60&            RXS-J00057-5007&    320.7098&    -65.4054&   0.0333&   24.1&   0&  13.69$\pm$ 0.01&    3.1$\pm$  1.5&  13.70$\pm$ 0.02&    6.1$\pm$  2.9&  2EUVE J0005-50.0 \\
 61&                 PG1216+069&    281.0714&     68.1432&   0.3313&   22.9&   0&  13.51$\pm$ 0.02&  -12.0$\pm$  2.2&  13.52$\pm$ 0.04&   -3.8$\pm$  4.5&  PG 1216+069 \\
 62&                  4C--01.61&     91.6647&    -60.3620&   0.1740&   22.6&   0&  13.38$\pm$ 0.02&    1.5$\pm$  3.0&  13.37$\pm$ 0.04&    2.0$\pm$  5.7&  4C -01.61 \\
 63&              QSO-B1215+303&    188.8752&     82.0529&   0.2370&   22.2&   0&  13.34$\pm$ 0.03&  -29.0$\pm$  3.8&  13.38$\pm$ 0.05&  -21.8$\pm$  6.6&  7C 121521.39+302340.00 \\
 64&                HE2258-5524&    330.7246&    -55.6698&   0.1400&   21.8&   0&  13.36$\pm$ 0.03&    8.1$\pm$  3.2&  13.39$\pm$ 0.05&   12.5$\pm$  6.3&  6dFGS gJ230152.0-550831 \\
 65&                       IZW1&    123.7486&    -50.1750&   0.0600&   21.4&   0&  13.40$\pm$ 0.02&  -17.0$\pm$  3.2&  13.42$\pm$ 0.05&  -22.5$\pm$  6.4&  Mrk 1502 \\
 \hline
 \hline
 \end{tabular}
 \end{center}
 \end{table}
 \end{turnpage}

\begin{turnpage}
\begin{table}
\renewcommand{\arraystretch}{0.9}
\caption*{Table 1 continued}
\begin{center}
\begin{tabular}{llrrrrrrrrrl}
\hline
\hline
No.&                   HSLA-ID&      Glon    &     Glat   &        $z$&   S/N& $Q$ &  log$N_{1393}$  &  $v_{c, 1393}$  &  log$N_{1402}$  &  $v_{c, 1402}$  & Simbad ID\\
   &                          &($\degree$) &($\degree$)&         & & &                 &  (km s$^{-1}$)    &                 &  (km s$^{-1}$)    &          \\
(0)&                   (1)    &      (2)   &     (3)   &   (4)   &   (5)  &(6)&            (7)  &  (8)            &         (9)     &  (10)           & (11)     \\
\hline
 66&    SDSSJ232259.98-005359.2&     80.2739&    -56.2638&   0.1505&   21.4&   0&  13.45$\pm$ 0.03&   14.1$\pm$  3.5&  13.49$\pm$ 0.05&    6.2$\pm$  6.1&  2MASX J23230004-0053588 \\
 67&                     MRK290&     91.4892&     47.9534&   0.0302&   21.3&   0&  13.49$\pm$ 0.02&  -29.2$\pm$  3.0&  13.50$\pm$ 0.04&  -18.6$\pm$  5.2&  Mrk 290 \\
 68&            RXS-J21388-3828&      4.5094&    -48.4648&   0.1830&   20.9&   0&  13.64$\pm$ 0.02&   19.3$\pm$  2.2&  13.62$\pm$ 0.03&   26.0$\pm$  4.2&  2MASX J21384992-3828403 \\
 69&                 PG1222+216&    255.0731&     81.6598&   0.4338&   20.7&   0&  13.63$\pm$ 0.02&   -7.2$\pm$  2.0&  13.66$\pm$ 0.03&    2.5$\pm$  3.6&  4C 21.35 \\
 70&                   MRK-1044&    179.6944&    -60.4774&   0.0161&   20.6&   0&  13.45$\pm$ 0.02&  -30.9$\pm$  3.4&  13.49$\pm$ 0.04&  -34.3$\pm$  6.2&  Mrk 1044 \\
 71&                  Q0439-433&    247.9801&    -41.3828&   0.5937&   19.7&   0&  13.67$\pm$ 0.02&   -1.5$\pm$  1.9&  13.69$\pm$ 0.03&    0.5$\pm$  3.5&  QSO J0441-4313 \\
 72&    2MASX-J10053271-2417161&    261.2303&     24.8511&   0.1540&   19.2&   0&  13.53$\pm$ 0.04&   29.5$\pm$  5.4&  13.55$\pm$ 0.03&   36.6$\pm$  5.1&  2MASX J10053271-2417161 \\
 73&             QSO-B1440+3539&     59.2376&     65.0346&   0.0770&   18.8&   0&  13.52$\pm$ 0.03&  -51.4$\pm$  4.9&  13.56$\pm$ 0.04&  -48.4$\pm$  6.5&  Mrk 478 \\
 74&                     RBS144&    299.4837&    -65.8362&   0.0628&   18.7&   0&  13.49$\pm$ 0.02&    1.1$\pm$  2.9&  13.51$\pm$ 0.05&   29.8$\pm$  6.5&  2MASX J01002713-5113544 \\
 75&                 PG1112+431&    167.8870&     64.9397&   0.3010&   18.3&   0&  13.54$\pm$ 0.02&  -36.5$\pm$  3.6&  13.54$\pm$ 0.04&  -41.6$\pm$  6.8&  QSO B1112+4306 \\
 76&                 PG1001+291&    200.0834&     53.2070&   0.3297&   18.3&   0&  13.64$\pm$ 0.02&  -27.3$\pm$  3.0&  13.64$\pm$ 0.03&  -27.6$\pm$  4.9&  Ton 28 \\
 77&                    RBS2055&    106.6724&    -34.6586&   0.0400&   18.3&   0&  13.73$\pm$ 0.02&  -12.3$\pm$  1.9&  13.76$\pm$ 0.03&  -10.1$\pm$  3.3&  2MASX J23515277+2619325 \\
 78&             QSO-B1617+1731&     32.8943&     41.0760&   0.1120&   18.2&   0&  13.82$\pm$ 0.02&  -22.7$\pm$  2.0&  13.85$\pm$ 0.03&  -20.5$\pm$  3.3&  Mrk 877 \\
 79&                HS1831+5338&     82.5864&     24.2127&   0.0450&   18.2&   0&  13.69$\pm$ 0.02&  -24.1$\pm$  2.6&  13.68$\pm$ 0.03&  -20.1$\pm$  4.4&  2MASX J18324966+5340219 \\
 80&                    MRK1148&    123.0934&    -45.4387&   0.0640&   18.1&   0&  13.41$\pm$ 0.03&   -5.9$\pm$  3.5&  13.46$\pm$ 0.05&  -18.3$\pm$  7.0&  Mrk 1148 \\
 81&                   B0120-28&    227.7753&    -82.9332&   0.4360&   18.0&   0&  13.60$\pm$ 0.02&  -12.3$\pm$  2.8&  13.64$\pm$ 0.04&   -3.1$\pm$  4.4&  QSO B0120-28 \\
 82&                   NGC-3516&    133.2357&     42.4028&      N/A&   17.8&   0&  13.31$\pm$ 0.04&  -50.9$\pm$  7.1&  13.36$\pm$ 0.06&  -50.5$\pm$  9.8&  [MWP92] A \\
 83&                HS1102+3441&    188.5631&     66.2187&   0.5083&   17.7&   0&  13.68$\pm$ 0.02&  -24.3$\pm$  2.7&  13.70$\pm$ 0.03&  -27.0$\pm$  4.4&  Ton 1329 \\
 84&                PG-1338+416&     90.5870&     72.4836&   1.2144&   17.7&   0&  13.41$\pm$ 0.03&  -41.8$\pm$  5.3&  13.42$\pm$ 0.06&  -28.8$\pm$  8.5&  QSO J1341+4123 \\
 85&                     TON236&     44.4335&     55.3506&   0.4474&   17.6&   0&  13.72$\pm$ 0.02&  -44.5$\pm$  3.9&  13.70$\pm$ 0.03&  -36.4$\pm$  5.2&  Ton 236 \\
 86&                    RBS1892&    345.8980&    -58.3672&   0.1980&   17.5&   0&  13.52$\pm$ 0.03&  -10.4$\pm$  3.2&  13.56$\pm$ 0.04&   -2.4$\pm$  5.3&  2MASS J22452029-4652116 \\
 87&              HB89-0232-042&    174.4627&    -56.1555&   1.4250&   17.4&   0&  13.38$\pm$ 0.03&   -8.0$\pm$  4.1&  13.41$\pm$ 0.06&    1.9$\pm$  7.8&  [RKV2003]--\footnote{[RKV2003] QSO J0235-0402 abs 1.425} \\
 88&                       3C48&    133.9626&    -28.7195&   0.3690&   17.3&   0&  13.51$\pm$ 0.03&  -23.1$\pm$  3.7&  13.48$\pm$ 0.05&  -13.2$\pm$  6.8&  3C 48 \\
 89&                    RBS1666&    358.7326&    -31.0031&   0.0796&   17.2&   0&  13.89$\pm$ 0.02&   28.0$\pm$  2.4&  13.89$\pm$ 0.02&   26.8$\pm$  3.3&  6dFGS gJ200553.0-413442 \\
 90&      2XMM-J100420.0+051300&    234.1610&     44.6222&   0.1600&   17.0&   0&  13.54$\pm$ 0.03&    6.4$\pm$  3.4&  13.54$\pm$ 0.05&    8.7$\pm$  6.3&  2MASS J10042013+0513004 \\
 91&                 PG1121+422&    167.2578&     66.8587&   0.2249&   16.9&   0&  13.60$\pm$ 0.02&  -38.9$\pm$  3.9&  13.64$\pm$ 0.04&  -23.7$\pm$  5.4&  2MASSI J1124392+420144 \\
 92&    SDSSJ094733.21+100508.7&    225.3726&     43.5395&   0.1390&   16.7&   0&  13.59$\pm$ 0.02&   -9.0$\pm$  2.7&  13.60$\pm$ 0.05&  -21.6$\pm$  6.8&  2MASX J09473320+1005093 \\
 93&             RXSJ00437+3725&    121.2330&    -25.4241&   0.0800&   16.3&   0&  13.50$\pm$ 0.03&  -25.4$\pm$  4.1&  13.51$\pm$ 0.05&  -31.8$\pm$  7.7&  LEDA 2100384 \\
 94&                  ZW535.012&    120.1743&    -17.1260&   0.0477&   16.1&   0&  13.30$\pm$ 0.05&  -16.3$\pm$  5.9&  13.30$\pm$ 0.08&  -20.0$\pm$ 11.2&  2MASX J00362092+4539532 \\
 95&             RXJ0439.6-5311&    261.2159&    -40.9266&   0.2430&   16.0&   0&  13.48$\pm$ 0.03&   -5.9$\pm$  3.6&  13.50$\pm$ 0.06&   -3.3$\pm$  7.0&  [VV98] J043938.7-531131 \\
 96&    SDSSJ141542.90+163413.8&      8.8491&     67.8261&   0.7426&   15.9&   0&  13.84$\pm$ 0.02&  -37.4$\pm$  3.1&  13.86$\pm$ 0.03&  -35.0$\pm$  4.3&  QSO J1415+1634 \\
 97&                    RBS2000&    350.1951&    -67.5844&   0.1736&   15.6&   0&  13.60$\pm$ 0.03&  -12.1$\pm$  3.1&  13.63$\pm$ 0.05&  -17.3$\pm$  5.8&  2FHL J2324.7-4041 \\
 98&                    PHL2525&     80.6832&    -71.3172&   0.2000&   15.4&   0&  13.68$\pm$ 0.02&  -19.7$\pm$  3.0&  13.71$\pm$ 0.04&  -32.1$\pm$  5.5&  GD 1332 \\
 99&             RXSJ00508+3536&    122.7965&    -27.2597&   0.0580&   15.3&   0&  13.47$\pm$ 0.03&  -29.2$\pm$  4.8&  13.46$\pm$ 0.06&  -25.2$\pm$  8.7&  2MASX J00505076+3536430 \\
100&    SDSSJ110307.57+291230.0&    201.5586&     66.0912&   0.3655&   15.0&   0&  13.52$\pm$ 0.04&   -8.7$\pm$  3.9&  13.50$\pm$ 0.05&  -14.4$\pm$  6.4&  [VV96] J110307.6+291230 \\
\hline
\hline
\end{tabular}
\end{center}
\end{table}
\end{turnpage}

\begin{turnpage}
\begin{table}
\renewcommand{\arraystretch}{0.9}
\caption*{Table 1 continued}
\begin{center}
\begin{tabular}{llrrrrrrrrrl}
\hline
\hline
No.&                   HSLA-ID&      Glon    &     Glat   &        $z$&   S/N& $Q$ &  log$N_{1393}$  &  $v_{c, 1393}$  &  log$N_{1402}$  &  $v_{c, 1402}$  & Simbad ID\\
   &                          &($\degree$) &($\degree$)&         & & &                 &  (km s$^{-1}$)    &                 &  (km s$^{-1}$)    &          \\
(0)&                   (1)    &      (2)   &     (3)   &   (4)   &   (5)  &(6)&            (7)  &  (8)            &         (9)     &  (10)           & (11)     \\
\hline
101&                  Q1545+210&     33.8983&     49.4574&   0.2643&   14.9&   0&  13.84$\pm$ 0.02&  -24.3$\pm$  2.7&  13.86$\pm$ 0.03&  -23.7$\pm$  4.1&  2MASX J15474353+2052167 \\
102&                 CAL-F-COPY&    277.1769&    -35.4212&   0.0640&   14.8&   0&  13.52$\pm$ 0.03&   14.1$\pm$  4.0&  13.54$\pm$ 0.06&   14.7$\pm$  7.4&  6dFGS gJ050304.0-663346 \\
103&                    TON1187&    188.3286&     55.3777&   0.0789&   14.8&   0&  13.48$\pm$ 0.03&  -25.7$\pm$  4.7&  13.49$\pm$ 0.06&  -20.5$\pm$  7.9&  Ton 1187 \\
104&             FBQSJ1010+3003&    198.4272&     54.6251&   0.2560&   14.7&   0&  13.61$\pm$ 0.03&  -29.9$\pm$  4.4&  13.63$\pm$ 0.05&  -34.8$\pm$  7.0&  Ton 488 \\
105&                  Q0349-146&    205.4845&    -46.3250&   0.6170&   14.6&   0&  13.44$\pm$ 0.04&  -13.2$\pm$  4.5&  13.49$\pm$ 0.06&   -1.0$\pm$  7.8&  2MASS J03512857-1429082 \\
106&              PMNJ1103-2329&    273.1898&     33.0793&   0.1860&   14.5&   0&  13.36$\pm$ 0.04&   37.3$\pm$  7.2&  13.41$\pm$ 0.07&   24.5$\pm$ 11.0&  2FHL J1104.0-2331 \\
107&        HB89-0107-025-NED05&    134.0294&    -64.7793&   0.9560&   14.4&   0&  13.48$\pm$ 0.04&  -14.9$\pm$  4.5&  13.49$\pm$ 0.07&  -19.5$\pm$  9.2&  QSO J0110-0218 \\
108&                     RBS563&    272.2534&    -39.2259&   0.0690&   14.2&   0&  13.23$\pm$ 0.05&  -10.4$\pm$  6.6&  13.29$\pm$ 0.13&    7.1$\pm$ 16.7&  2MASX J04382919-6147587 \\
109&                HE2259-5524&    330.6372&    -55.7230&   0.8510&   14.2&   0&  13.46$\pm$ 0.04&   10.2$\pm$  4.6&  13.45$\pm$ 0.07&   11.1$\pm$  9.3&  [VV2000] J230222.5-550827 \\
110&              QSO-B1229+204&    269.4424&     81.7396&   0.0637&   14.1&   0&  13.61$\pm$ 0.03&  -18.3$\pm$  3.4&  13.63$\pm$ 0.05&  -30.5$\pm$  7.2&  Mrk 771 \\
111&    SDSSJ092909.79+464424.0&    172.5843&     46.0101&   0.2400&   14.1&   0&  13.28$\pm$ 0.06&  -33.1$\pm$  8.6&  13.36$\pm$ 0.08&   -6.7$\pm$ 11.0&  2MASS J09290979+4644240 \\
112&   SDSS-J141309.14+092011.2&    354.0972&     63.7588&   0.4596&   14.0&   0&  13.61$\pm$ 0.03&   -2.0$\pm$  3.2&  13.63$\pm$ 0.05&   -4.1$\pm$  5.7&  2MASS J14130915+0920109 \\
113&              PMNJ2345-1555&     65.6728&    -70.9851&   0.6210&   13.7&   0&  13.62$\pm$ 0.03&  -12.2$\pm$  3.5&  13.67$\pm$ 0.05&  -14.6$\pm$  6.1&  3FGL J2345.2-1554 \\
114&                    4C25.01&    114.0722&    -36.2766&   0.2840&   13.5&   0&  13.69$\pm$ 0.03&  -15.1$\pm$  3.0&  13.74$\pm$ 0.04&  -12.6$\pm$  5.5&  2MASS J00193977+2602522 \\
115&                 PG0923+201&    210.1766&     42.6498&   0.1922&   13.5&   0&  13.25$\pm$ 0.06&  -19.1$\pm$  8.5&  13.29$\pm$ 0.10&  -26.9$\pm$ 13.9&  Ton 1057 \\
116&                HE0435-5304&    261.0243&    -41.3745&   1.2310&   13.3&   0&  13.48$\pm$ 0.04&   -3.5$\pm$  4.3&  13.56$\pm$ 0.07&  -12.5$\pm$  8.6&  [VV2000] J043650.8-525849 \\
117&              QSO-B2356-309&     12.8390&    -78.0354&   0.1654&   13.2&   0&  13.39$\pm$ 0.05&    3.9$\pm$  5.5&  13.45$\pm$ 0.08&   -5.4$\pm$  9.7&  QSO B2356-309 \\
118&                   4C-13.41&    225.1177&     49.1204&   0.2410&   13.1&   0&  13.66$\pm$ 0.03&   -2.4$\pm$  3.3&  13.67$\pm$ 0.05&    8.3$\pm$  6.2&  2XMM J100726.0+124856 \\
119&                 PG0832+251&    199.4909&     33.1462&   0.3299&   12.8&   0&  13.43$\pm$ 0.04&  -13.3$\pm$  5.4&  13.42$\pm$ 0.08&   -2.1$\pm$  9.9&  QSO B0832+2510 \\
120&           IRAS-F04250-5718&    266.9871&    -41.9964&   0.1040&   73.9&   1&  13.58$\pm$ 0.01&  -12.9$\pm$  0.6&  13.67$\pm$ 0.01&  -11.3$\pm$  1.0&  LB 1727 \\
121&                PKS2005-489&    350.3731&    -32.6008&   0.0710&   21.0&   1&  14.00$\pm$ 0.01&   40.1$\pm$  2.5&  14.06$\pm$ 0.02&   42.7$\pm$  2.5&  QSO B2005-489 \\
122&                 PG1211+143&    267.5518&     74.3150&   0.0810&   18.6&   1&  13.58$\pm$ 0.02&  -10.8$\pm$  2.5&  13.76$\pm$ 0.03&   -4.4$\pm$  3.0&  2E 2620 \\
123&                 PKS1136-13&    277.5252&     45.4316&   0.5565&   18.5&   1&  13.83$\pm$ 0.02&   28.2$\pm$  2.3&  13.93$\pm$ 0.02&   29.6$\pm$  3.0&  NAME Crt A \\
124&                PKS0558-504&    257.9623&    -28.5693&   0.1370&   18.5&   1&  13.70$\pm$ 0.02&    9.6$\pm$  2.0&  13.81$\pm$ 0.02&    7.0$\pm$  2.9&  QSO B0558-5026 \\
125&                 PG1435-067&    343.9784&     47.2062&   0.1290&   18.3&   1&  13.61$\pm$ 0.02&  -36.4$\pm$  3.3&  13.70$\pm$ 0.03&  -36.3$\pm$  4.9&  2E 3305 \\
126&    2MASS-J14312586+2442203&     32.4155&     67.4228&   0.4070&   17.2&   1&  13.78$\pm$ 0.02&  -40.9$\pm$  3.2&  13.87$\pm$ 0.03&  -43.8$\pm$  4.5&  2MASSI J1431258+244220 \\
127&                    RBS2005&     97.9384&    -36.7987&   0.1200&   16.7&   1&  13.75$\pm$ 0.02&    1.3$\pm$  1.8&  13.88$\pm$ 0.03&    1.3$\pm$  3.0&  2MASS J23255424+2153140 \\
128&                     MRK841&     11.2089&     54.6320&   0.0364&   15.8&   1&  13.88$\pm$ 0.02&  -41.2$\pm$  3.6&  14.03$\pm$ 0.03&  -42.4$\pm$  4.0&  Mrk 841 \\
129&                     PDS456&     10.3923&     11.1632&   0.1850&   14.2&   1&  14.00$\pm$ 0.02&   -7.5$\pm$  1.9&  14.03$\pm$ 0.03&    0.6$\pm$  2.4&  QSO B1725-142 \\
130&             RXJ2139.7+0246&     58.0875&    -35.0060&   0.2600&   12.5&   1&  13.75$\pm$ 0.03&    4.8$\pm$  2.9&  13.84$\pm$ 0.04&    1.9$\pm$  4.1&  2MASS J21394418+0246052 \\
131&                      3C273&    289.9509&     64.3600&   0.1580&  134.4&  -1&  14.21$\pm$ 0.01&    3.0$\pm$  3.0&  13.82$\pm$ 0.00&  -19.9$\pm$-19.9&  3C 273 \\
132&                PKS0405-123&    204.9271&    -41.7563&   0.5726&   77.0&  -1&  13.46$\pm$ 0.01&  -20.3$\pm$-20.3&  13.25$\pm$ 0.02&  -14.9$\pm$-14.9&  PKS 0405-123 \\
\hline
\hline
\end{tabular}
\end{center}
\end{table}
\end{turnpage}


\end{document}